\PassOptionsToPackage{numbers,sort&compress}{natbib}
\documentclass[preprintnumbers,amsmath,amssymb,prd, notitlepage,nofootinbib, superscriptaddress]{revtex4}
\usepackage{amsfonts,amssymb,amsmath}
\usepackage{gensymb}
\usepackage{color}
\usepackage{graphicx}
\usepackage{multirow}
\usepackage[utf8]{inputenc}
\usepackage{aas_macros}

\graphicspath{{./}}
\usepackage[normalem]{ulem}

\newcommand{\greaterthanapprox}{\mathrel{\vcenter{
  \offinterlineskip\halign{\hfil$##$\cr
    >\cr\noalign{\kern2pt}\sim\cr\noalign{\kern-2pt}}}}}
    
    \newcommand{\lessthanapprox}{\mathrel{\vcenter{
  \offinterlineskip\halign{\hfil$##$\cr
    <\cr\noalign{\kern2pt}\sim\cr\noalign{\kern-2pt}}}}}
    
\newcommand{\lb}{\left(}
\newcommand{\rb}{\right)}

\newcommand{\Planck}{{\it Planck}~}

\newcommand{\lsb}{\left[}
\newcommand{\rsb}{\right]}
    
\newcommand{\ti}{\textit}

\newcommand{\be}{\begin{equation}}        
\newcommand{\ee}{\end{equation}}
\newcommand{\ba}[1]{\begin{align}
#1
\end{align}}

\DeclareMathOperator{\Tr}{Tr}

\begin{document}

\title{Improving models of the cosmic infrared background using CMB lensing mass maps}

\author{Fiona McCarthy}
\email{fmccarthy@perimeterinstitute.ca}

\affiliation{Perimeter Institute for Theoretical Physics, Waterloo, Ontario, N2L 2Y5, Canada}
\affiliation{Department of Physics and Astronomy, University of Waterloo, Waterloo, Ontario, Canada, N2L 3G1}

\author{Mathew S. Madhavacheril}
\affiliation{Perimeter Institute for Theoretical Physics, Waterloo, Ontario, N2L 2Y5, Canada}

\date{\today}

\begin{abstract}
The cosmic infrared background (CIB) sourced by infrared emission from dusty star-forming galaxies is a valuable source of information on the star formation history of the Universe. In measurements of the millimeter sky at frequencies higher than $\sim 300$ GHz, the CIB and thermal emission from Galactic dust dominate.  Insufficient understanding of the CIB contribution at lower frequencies can hinder efforts to measure the kinetic Sunyaev-Zeldovich spectrum on small scales as well as new physics that affects the damping tail of the cosmic microwave background (CMB). The {\it Planck} satellite has measured with high fidelity the CIB at 217, 353, 545 and 857 GHz. On very large scales, this measurement is limited by our ability to separate the CIB from Galactic dust, but on intermediate scales, the measurements are limited by sample variance in the underlying matter field traced by the CIB. We show how significant improvements (20-100\%) can be obtained on parameters of star formation models by cross-correlating the CIB (as measured from existing {\it Planck} maps or upcoming CCAT-prime maps)  with upcoming mass maps inferred from gravitational lensing of the CMB. This improvement comes from improved knowledge of the redshift distribution of star-forming galaxies as well as through the use of the unbiased matter density inferred from CMB lensing mass maps to cancel the sample variance in the CIB field. We also find that further improvements can be obtained on CIB model parameters if the cross-correlation of the CIB with CMB lensing is measured over a wider area while restricting the more challenging CIB auto-spectrum measurement to the cleanest 5\% of the sky. 
 \end{abstract}
\maketitle

\section{Introduction}

\noindent

Star forming galaxies contain particles of dust that absorb ultraviolet light and emit thermally in the infrared (IR). This IR emission sources the Cosmic Infrared Background (CIB), a diffuse, unresolved background that traces star-forming galaxies. Dust content in galaxies is correlated with the star formation rate, and the CIB emissivity peaks at around redshift $z\sim2$ where the star formation rate is high.  Anisotropies in the CIB trace anisotropies in the star-forming galaxy distribution \cite{2001ApJ...550....7K} and give insight into the physics of star formation. The CIB, and the CIB anisotropies, have been detected at numerous wavelengths \cite{Addison_2012} by IRIS \cite{2005ApJS..157..302M}, Herschel \cite{2013ApJ...772...77V}, SPT \cite{2010ApJ...718..632H}, \Planck \cite{2011A&A...536A..18P, 2014A&A...571A..30P} and ACT \cite{2013JCAP...07..025D,Choi:2020ccd}. Various theoretically motivated parametric models have been fit to the  data measuring the CIB (e.g. \cite{Wu:2016vpb,Maniyar:2018xfk,2014A&A...571A..30P,2013ApJ...772...77V,Maniyar:2020tzw}). Improving these models is not just useful for understanding star formation history itself, but also because the CIB appears as a foreground to the cosmic microwave background (CMB) at lower frequencies. 

\begin{figure}[t]
\begin{center}
\includegraphics[width=0.45\textwidth]{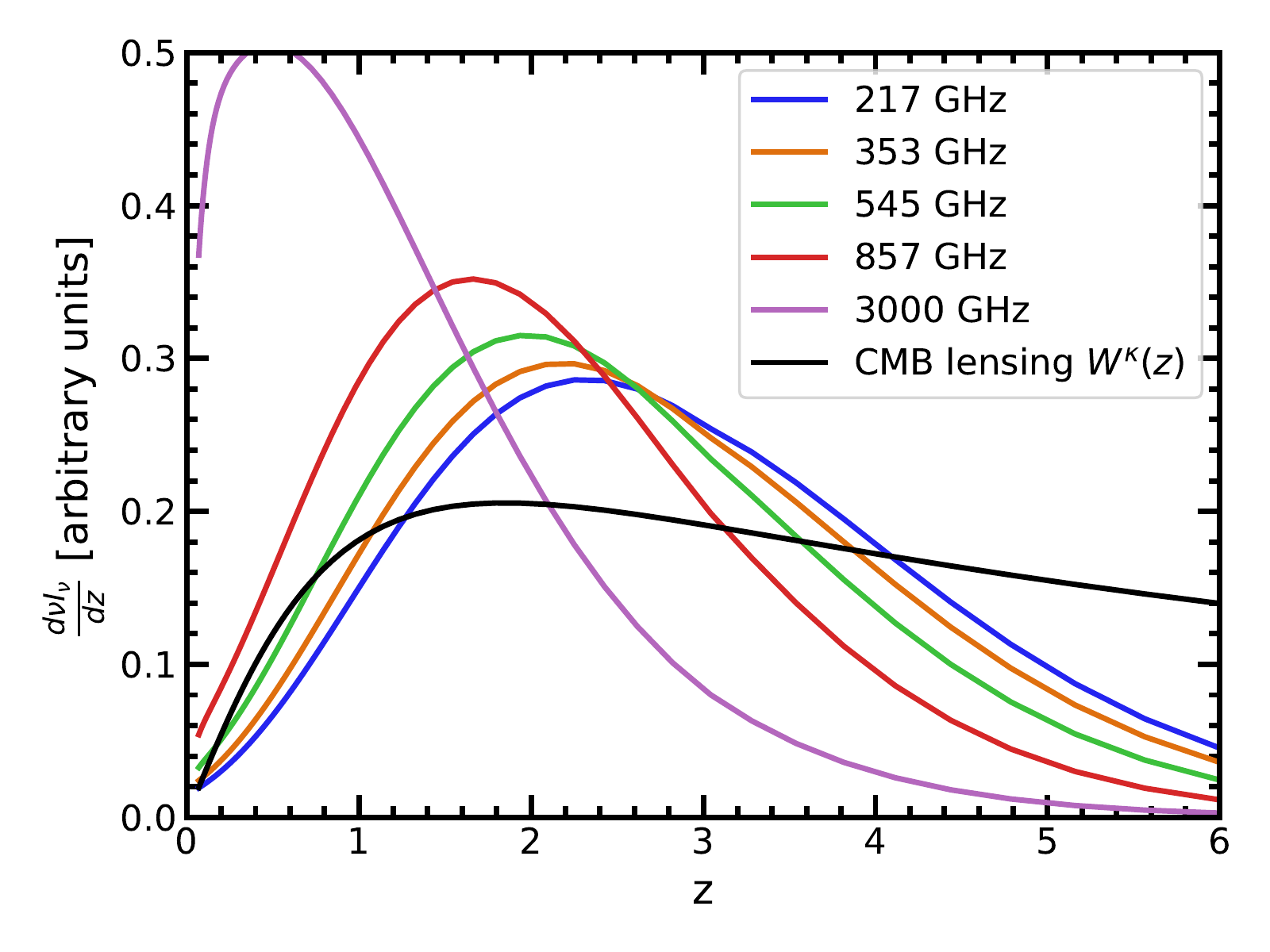}
\includegraphics[width=0.45\textwidth]{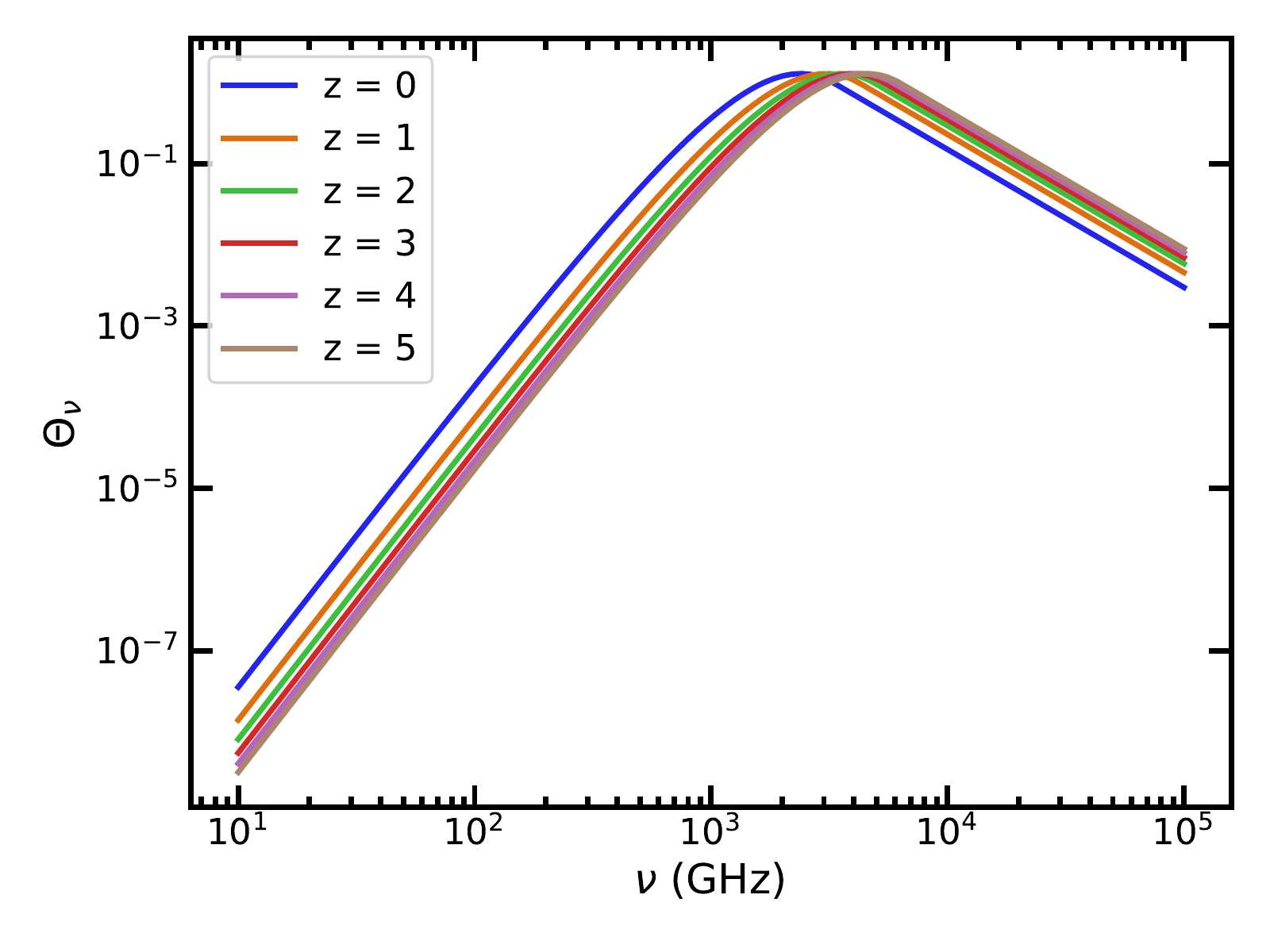}
\caption{{\it Left:} The redshift distribution of the CMB lensing kernel and the CIB. On the top is the redshift distribution of the CIB monopole $\frac{d \nu I _\nu}{dz}$, with all kernels normalised so that $\int_{0}^6 W(z)dz= 1$; the (similarly normalised) CMB lensing efficiency kernel $W^\kappa(z)$ is also shown in black. {\it Right:} The parametric CIB SED at fixed redshifts, normalised as in Equation \eqref{SED_parametric}}\label{fig:redshift_distribution}
\end{center}
\end{figure}

Gravitational lensing of the CMB offers an unbiased probe of the total matter content of the universe. While the CMB is sourced at very high redshift $z\sim1100$, it is well known that the CMB we detect has been lensed by intervening matter \cite{Lewis:2006fu}. The lensing kernel of the CMB peaks at $z\sim2$, close to where the CIB intensity density peaks (see Figure \ref{fig:redshift_distribution}), and as the galaxies sourcing the CIB trace the dark matter primarily responsible for CMB lensing, it is expected (and confirmed empirically e.g. \cite{Ade:2013aro,vanEngelen:2014zlh,Darwish:2020fwf}) that the CMB lensing potential and the CIB are correlated. This high degree of correlation has been exploited for improving the science return from CMB experiments, e.g., by using the CIB as (or as part of) a template \cite{2017PhRvD..96l3511Y} for the lensing signal itself, allowing one to undo the effect of the lensing signal. Delensing the CMB in this way \cite{2016PhRvL.117o1102L,GreenDelensing} allows us to more clearly reveal underlying cosmological signals of interest, e.g. B-modes from primordial gravitational waves \cite{2020arXiv200812619T} or new relativistic species \cite{GreenDelensing}.

In this work, we explore the potential for obtaining improved models of the CIB from cross-correlations of the CMB lensing signal with existing measurements of the CIB from \Planck and future high-resolution measurements from CCAT-prime \cite{1807.04354,2020JLTP..199.1089C}. Such improvements will enhance our understanding of high-redshift star formation and will relax degeneracies encountered in the damping tail of the CMB temperature power spectrum. As an example of the latter, inferences of the amplitude of the kinetic Sunyaev-Zeldovich effect (e.g. \cite{Reichardt:2020jrr}) can be affected by model bias in the CIB contribution.

The CMB lensing potential can be reconstructed from statistical anisotropies in the CMB \cite{Okamoto:2003zw}: the \Planck collaboration has reconstructed the lensing potential on about $70\%$ percent of the sky \cite{1807.06210} with signal-to-noise per mode close to unity near the peak of the power spectrum, but otherwise generally noise-dominated. High-resolution ground-based CMB experiments like ACT and SPT are now making CMB maps that are significantly signal dominated over a larger range of scales (albeit currently on small fractions of the sky)\cite{Darwish:2020fwf,1905.05777}. 

Over the next decade however, the CMB lensing potential will be imaged with high fidelity to even higher $L$ than at present over large fractions of the sky \cite{Ade:2018sbj,Abazajian:2016yjj}. We also expect improvements in CIB measurements in coming years, with experiments such as CCAT-prime \cite{1807.04354,2020JLTP..199.1089C}, and Simons Observatory in its highest frequency channels \cite{Ade:2018sbj}, measuring the small-scale CIB to higher accuracy. The CMB lensing / CIB cross correlation has already been used to fit large-scale (linear) models of the CIB \cite{Ade:2013aro,Maniyar:2018xfk,Cao:2019hmt}; our work here forecasts the improved parameter constraints on the parametric halo model for the CIB introduced in  \cite{2012MNRAS.421.2832S}, which was fit to \Planck+IRIS CIB power spectrum data in \cite{2014A&A...571A..30P}. While the CMB lensing potential does not depend on the CIB model parameters, we expect improvements due to the cross-correlation depending on the redshift distribution of the CIB as well as due to sample-variance cancellation (see e.g. \cite{Seljak:2008xr}), where measuring the CMB lensing potential on the same patch of sky as the CIB intensity can afford improvements in the CIB model due to their high correlation.

CIB models have previously been cross-correlated with CMB lensing maps to infer CIB model parameters \cite{Ade:2013aro,Maniyar:2018xfk,Cao:2019hmt} and also with other external large-scale-structure probes such as the Sloan Digital Sky Survey (SDSS) galaxies \cite{Serra:2014pva}. In this work we quantify the potential improvements of employing such external cross-correlations in particular as we get access to better CMB lensing data.

In all our calculations, we use the cosmology of \cite{Ade:2013zuv}: $\left\{\Omega_m, \Omega_\Lambda, \Omega_bh^2, 10^9A_s, h, n_s\right\}=\{0.3175, 0.6825, 0.022068, 2.2,$\\
$ 0.6711, 0.9624\}$. In our halo model, we use the halo bias, halo mass function, and subhalo mass function of Tinker \cite{2010ApJ...724..878T, 2010ApJ...719...88T}, and we use Navarro--Frenk--White  (NFW) \cite{1996ApJ...462..563N} halo profiles. We explore constraints from subsets of measurements of the CIB made by \Planck at 217, 353, 545, 857 GHz, IRIS at 3000 GHz \cite{2005ApJS..157..302M} and future CCAT-prime measurements \cite{1807.04354,2020JLTP..199.1089C} at \{220,  280,  350,  410,  850\} GHz , with improvements from CMB lensing measured by \Planck or future Simons Observatory-like \cite{Ade:2018sbj} and CMB-S4-like \cite{Abazajian:2016yjj} experiments.  For  all CMB lensing fields, we impose a maximum multipole of $L=1000$; for \Planck+IRIS CIB forecasts we use a multipole range of $186\le L\le2649$, and for CCAT-prime we include $L\le10000$. Our baseline forecasts are on 2240 square degrees of sky, with the 3000 GHz IRIS field only included on 183 square degrees, mimicking the analysis of \cite{2014A&A...571A..30P}. CMB lensing reconstruction maps are assumed to have full overlap with these 2240 square degrees, but we also explore the possibility of including the CIB / CMB lensing cross-correlations over larger fractions of the sky without including the CIB auto-spectrum (which may have larger systematics due to Galactic dust).

The paper is organised as follows. In Sec. \ref{sec:halo_model} we review the theory of the CIB power spectrum within the halo model, and we present the specific parametric model we consider in Sec. \ref{sec:parametric_CIB}. This formalism follows very closely what has already been considered in e.g. \cite{2012MNRAS.421.2832S,Reischke:2019ikn,2013ApJ...772...77V,2014A&A...571A..30P}, which we reproduce here in detail. In Sec. \ref{sec:fisher_forecasts} we introduce our Fisher forecast formalism and present the experimental configurations we consider. In Sec. \ref{sec:fcresults} we present the results of our forecast on parameter constraints. We also discuss in Sec. \ref{sec:sfr} the improvements in constraints of the star formation rate. We discuss our results and conclude in Sec. \ref{sec:conclusion}.

\section{The CIB-CIB and CIB-CMB lensing power spectra within the halo model}\label{sec:halo_model}

The halo model (see e.g. \cite{Cooray:2002dia} for a review) has been widely used to model the distribution of matter and galaxies in discrete `halos', which form from the collapse of initially overdense regions and evolve non-linearly with gravity. Correlations are categorised either as large-scale correlations, between two separate halos (a `2-halo' term); or small-scale correlations within a single halo (a `1-halo' term). 
 In \cite{2012MNRAS.421.2832S} a halo model prescription was presented for the CIB, wherein  the CIB power spectrum is modelled as an emissivity-weighted version of the galaxy power spectrum. In this section we review the CIB emissivity model and its halo model prescription.

Within the halo model, all the matter in the universe is assumed to be in these halos which have undergone gravitational collapse; the halos can be described by their total mass $M$. All galaxies form within halos according to a halo occupation distribution (HOD); central galaxies at the centre of halos and satellite galaxies in sub-halos. The number of central galaxies hosted by a halo is assumed to be a function of $M$ (and, more generally, redshift $z$); similarly, the number of sub-halos of mass $M_s$ hosted by a halo is a function of $M$ and the subhalos follow the dark matter density profile of the halo $\rho(r)$, where $r$ is the distance from the centre of the halo (for a spherically symmetric halo). Due to the complex physics of galaxy formation, the properties of the galaxies (e.g. total stellar mass, luminosity) are not in general easy to calculate; however a simplifying assumption is that they are functions of their host halo mass (and redshift). Following this assumption,  a luminosity $L^{\rm gal}(M,z)$ is assigned to each galaxy according to its host halo (or sub-halo) mass $M$. Thus the luminosity of a \ti{halo} is the sum of the luminosities of all the galaxies it contains, including centrals and satellites. The total luminosity is weighted by a spectral energy distribution (SED) to find the luminosity density at frequency $\nu$ $L_\nu(M,z)$.

\subsection{CIB Emissivity Power Spectrum}
The CIB intensity density $I_\nu$ at frequency $\nu$ is a line-of-sight integral of emissivity density $j_\nu$ out to reionisation at $\chi_{re}$:
\be
I_\nu (\hat{\mathbf{ n}})= \int _0^{\chi_{re}}d\chi\, a(\chi )j_\nu\lb\chi, \hat{\mathbf{ n}}\rb.\label{Inu}
\ee
Writing the emissivity as a sum of the average emissivity $\bar j(z)$ and the fluctuation $\delta j (\hat{\mathbf{ n}} ,z)$ gives
\be
I_\nu (\hat{\mathbf{ n}})= \int _0^{\chi_{re}}d\chi\, a(\chi )\bar j_\nu(z)\lb1+\frac{\delta j _\nu(\hat{\mathbf{ n}} ,z)}{\bar j _\nu (z)}\rb\label{inu_expand}
\ee
which can be written as $I_\nu(\hat{\mathbf{ n}})=  I_\nu + \delta I_\nu (\hat{\mathbf{ n}})$, with $I_\nu$ the mean intensity at $\nu$.
The angular power spectrum of the CIB intensity anisotropies is defined as
\be
\left<\delta I_{L M}^\nu\delta I_{L^\prime M^\prime}^{\nu^\prime}\right>\equiv C_{L}^{\nu\nu^\prime}\delta_{LL^\prime}\delta_{MM^\prime}.
\ee
Employing the Limber approximation \cite{1953ApJ...117..134L} we can write the angular power spectrum as an integral over the three-dimensional emmisivity power spectrum
\be
C_{L}^{\nu\nu^\prime}=\int \frac{d\chi}{\chi^2}a^2\bar j_\nu(z) \bar j _{\nu^\prime}(z) P_{j}^{\nu\nu^\prime}\lb k=\frac{L}{\chi},z\rb
\ee
where  the power spectrum $ P_{j}^{\nu\nu^\prime}(k,z)$  of the three-dimensional fluctuations $\frac{\delta j_\nu}{j\nu}$ is given by
\be
\frac{\left<\delta j_\nu (\mathbf{ k},z)\delta j_{\nu^\prime} (\mathbf{ k}^\prime,z)\right>}{\bar j _\nu(z) \bar j _{\nu^\prime}(z)}\equiv \lb2\pi\rb^3 P_{j}^{\nu\nu^\prime}( k,z)\delta^3(\mathbf{ k} - \mathbf{ k}).
\ee

\subsubsection{Connecting to the Halo Model}

The emissivity density $j_\nu$ is an integral over luminosity density
\be
 j_\nu(z)=\int dL_{(1+z)\nu}\frac{d N}{dL_{(1+z)\nu}}\frac{L_{(1+z)\nu}}{4\pi}\label{j_def}
\ee
where $\frac{d N}{dL_\nu}$ is the luminosity function such that $ dL_\nu\frac{d N}{dL_\nu}$ gives the number density of galaxies with luminosity between $L_\nu$ and $L_\nu+dL_\nu$. The factor of $(1+z)$ in the frequency accounts for the redshift of the emitted radiation. Neglecting scatter between luminosity and halo mass, this can be written as an integral over halo mass $M$:
\be
 j_\nu(z)=\int dM\frac{d N}{dM}\frac{L_{(1+z)\nu}(M,z)}{4\pi}\label{j_M}
 \ee
where $\frac{d N}{dM}$ is the halo mass function. 

Equation \eqref{j_M} should be compared to the expression for the number density of galaxies within the halo model:
\be
\bar n_{\rm gal}=\int dM \frac{dN}{dM}N_{\rm gal}(M)\label{ngal}
\ee
the number density $\bar n_{\rm gal}$ is equal to an integral over the halo mass function, weighted by the number of galaxies hosted in a halo of mass $M$ $N_{\rm gal}(M)$.  We see from comparing Equations \eqref{j_M} and \eqref{ngal} that the power spectrum of $\bar j$ within the halo model can be arrived at from the galaxy power spectrum by replacing $N_{\rm gal}$ with the luminosity  $L_{(1+z)\nu}(M,z)/4\pi$.   As such, it is helpful to state  the 1- and 2-halo galaxy power spectra within the halo model. 

\subsubsection{Galaxy power spectra}

The galaxy power spectrum is defined as
\be
\frac{\left<\delta n_{\rm gal} (\mathbf{ k},z)\delta n_{\rm gal} (\mathbf{ k}^\prime,z)\right>}{\bar n_{\rm gal}(z)^2}\equiv \lb2\pi\rb^3 P^{gg}( k,z)\delta^3(\mathbf{ k} - \mathbf{ k}),
\ee
where the galaxy density is $n _{\rm gal}(z) = \bar n _{\rm gal}(z) + \delta n_{\rm gal}(z)$. 
On large scales, the halos are biased with respect to the underlying dark matter field, with a scale-dependent bias. The total galaxy power spectrum is thus often written on large scales as $P^{gg}(k,z)\sim b(z)^2 P_{\rm lin}(k,z)$, with $P_{\rm lin}(k,z)$ the linear matter power spectrum and $b$ the scale-dependent galaxy bias. However, halos of different masses are biased differently, and this expression is arrived at from
\begin{equation}
P_{gg}^{\rm 2-halo}(k,z) =\lb \int dM \frac{dN}{dM}\frac{N^{\rm cen}(M,z)+N^{\rm sat}(M,z)u(k,M,z)}{\bar n_{\rm gal}(z)} b(M,z) \rb ^2 P_{\rm lin}(k,z) \label{gal2halo}
\end{equation}
where $b(M,z)$ is the halo bias,  $u(k,M,z)$ is  the (normalised) Fourier transform of the halo density profile (equal to 1 on the large scales where the 2-halo term is dominant), and the galaxies are distinguished by whether they are central galaxies (which are hosted at the centre of the halo), and satellite galaxies: $N^{\rm cen}(M,z)$ is the number of central galaxies hosted by a halo of mass $M$ at $z$ and $N^{\rm sat}(M,z)$ is the number of satellite galaxies. For the 1-halo power spectrum, the substructure of the halos is more important; with both central-central and satellite-satellite correlations taken into account the 1-halo galaxy power spectrum is
\begin{widetext}
\be
P_{gg}^{\rm 1-halo}(k,z) = \int dM\frac{dN}{dM}\lb\frac{2N^{\rm sat}(M,z)N^{\rm cen}(M,z) u(k,m,z) + N^{\rm sat}(M,z)^2u(k,M,z)^2}{\bar n _{\rm gal}(z)^2}\rb.
\ee
\end{widetext}

\subsubsection{From galaxy power spectra to emissivity power spectra}

To write the 2-halo power emissivity power spectrum, we replace galaxy number $N_{\rm gal}(M,z)/\bar n_{\rm gal}(z)$ with $L_{\nu(1+z)}/4\pi\bar j_\nu (z)$. As such we have
\be
\bar j_\nu (z)\bar j_{\nu^\prime} (z)P_{j}^{\nu\nu^\prime}{}^{\rm 2-halo}(k,z)=D_\nu(z) D_{\nu^\prime}(z) P_{\rm lin}(k,z)\label{p2halo}
\ee
where $D_\nu(z)$ is the CIB bias weighted by $u(k,M,z)$ (without the $u(k,M,z)$ term the following integral would define the CIB bias)

\begin{equation}
D_\nu(z,k) \equiv\int dM \frac{dN}{dM}b(M,z)\lb\frac{L^{\mathrm{cen}}_{(1+z)\nu}(M,z)+L^{\mathrm{sat}}_{(1+z)\nu}(M,z)u(k,M,z)}{4\pi}\rb.
\end{equation}
where again the luminosity of a halo of mass $M$ $L_{(1+z)\nu}(M,z)$ comprises both the luminosity of a central galaxy and the luminosity of the satellite galaxies in subhalos:
\be
L_{(1+z)\nu}(M,z) = L^{\rm cen}_{(1+z)\nu}(M,z)+ L^{\rm sat}_{(1+z)\nu}(M,z).
\ee
The 2-halo term \eqref{p2halo} thus takes into account correlations between the galaxies of two different halos.

To write the 1-halo correlations, which include both central-satellite and satellite-satellite correlations \ti{within} a single halo, we write
\begin{widetext}
\begin{align}
&\bar j_\nu (z)\bar j_{\nu^\prime} (z)P_{j}{}^{\nu\nu^\prime}{}^{\rm 1-halo}(k,z)=\nonumber\\
&\int dM\frac{dN}{dM}\frac{1}{\lb4\pi\rb^2}\bigg{(} L_{(1+z)\nu}^{\rm cen} L_{(1+z)\nu^\prime}^{\rm sat}u(k,M,z)
+L_{(1+z)\nu^\prime}^{\rm cen} L_{(1+z)\nu}^{\rm sat}u(k,M,z)
+L_{(1+z)\nu}^{\rm sat} L_{(1+z)\nu^\prime}^{\rm sat}u^2(k,M,z)\bigg{)}.
\end{align}
\end{widetext}
\subsubsection{Central and Satellite luminosity}
As the luminosity of halos is sourced by the galaxies it is host to, the (central or satellite) luminosity must depend on the properties of the (central or satellite) galaxies. A simplifying assumption is that the luminosity of a galaxy depends on the mass of its host halo or subhalo in the same functional form for both central and satellite galaxies: $L^{\rm gal}(M^{\rm host},z)$, where $M^{\rm host}$ is the mass of the galaxy's host halo or host subhalo. As such, the central luminosity of a halo is simply
\be
 L^{\rm cen}_{(1+z)\nu}(M,z) = N^{\rm cen}(M,z)L^{\rm gal}_{(1+z)\nu}(M,z)
\ee 
where $N^{\rm cen}(M,z)$ is the number of central galaxies a halo of mass $M$ hosts. Then, using the assumption that satellite galaxy luminosity has the same dependence on host subhalo mass as central galaxy luminosity on host halo mass, the entire luminosity of a halo due to the satellite galaxies is an integral over the subhalos:
\be
L^{\rm sat}_\nu(M,z) = \int dM_s\frac{dN}{dM_s} L^{\rm gal}_{(1+z)\nu}(M_s,z)
\ee
where $\frac{dN}{dM_s}$ is the subhalo mass function; note that the number of satellite galaxies in a halo of mass $M$ is 
\be
N^{\rm sat}(M,z) = \int dM_s\frac{dN}{dM_s}.
\ee

\subsubsection{Poissonian term: shot noise}

There is also a shot noise component in the power spectrum, arising from the discrete nature of the sources. This is present in both the $\nu=\nu$ power spectra and the $\nu\nu^\prime$ spectra with $\nu\ne\nu^\prime$, as the same source can contribute to the intensity at different frequencies.

The shot noise is scale independent and is given by an integral over the flux density $S_\nu$ of all sources at frequency $\nu$  up to a cutoff frequency at which point sources are removed $S_{\rm cut}$:
\be
C_L=\int _0 ^{S_{\rm cut} } S_\nu^2 \frac{dN}{dS_\nu}dS_\nu.
\ee
$\frac{dN}{dS_\nu}$ is the distribution of flux densities such that $\frac{dN}{dS_\nu}\Delta S_\nu$ is the (angular) number density of sources with flux between $S_\nu$ and $S_\nu+\Delta S_\nu$.

\subsubsection{Point source removal and $S_{cut}$}

At current angular resolutions, the CIB is a diffuse, unresolved emission; however it is composed of discrete point sources---galaxies. If a single galaxy is bright enough, it can appear in a map as a point source and be removed. In each CIB map there is a (frequency-dependent) threshold flux density $S_{\rm cut}$ above which the point sources can be removed. Considering that flux $S_\nu$ can be expressed in terms of luminosity $L_{(1+z)\nu}$ by
\be
S_\nu = \frac{L_{(1+z)\nu}}{4\pi(1+z)\chi^2}\label{flux_to_lum},
\ee
a flux-cut is equivalent to a $z$-dependent luminosity cut, which should be imposed in the calculations of $C_L$.  We implement the flux cut by removing all halos with total luminosity greater than that corresponding to the flux limit in Equation \eqref{flux_to_lum} where $S^\nu$ is replaced by the flux cut of the experiment in question. 

\subsection{The CIB-CMB Lensing cross power spectrum}

The angular power spectrum of the CMB lensing potential $\phi$ is given in the Limber approximation by (see e.g.
 \cite{Lewis:2006fu})

\be
C_L^{\phi\phi}=\frac{4}{L^4}\int d\chi W^2_\kappa(\chi) P_{ mm}\lb k=\frac{L}{\chi},z\rb
\ee
with the lensing efficiency kernel $W_\kappa(\chi)$ given by
\be
W_\kappa(\chi)= \frac{3}{2}\lb\frac{H_0}{c}\rb^2 \frac{\Omega_m}{a}\chi \lb1-\frac{\chi}{\chi_S}\rb
\ee
where $\chi_S$ is the comoving distance of the source of the CMB at $z\sim1100$ and $P_{\rm mm}(k,z)$ is the matter power spectrum. The matter power spectrum can be computed within the halo model by 
\be
P_{mm}^{2h}(k,z)=\lb\int dM \frac{dn}{dM }\frac{M}{\rho_m}b(M,z)u(k,M,z) \rb^2 P_{\rm lin}(k,z)
\ee
(with $\rho_m$ the matter density today), where the ``dark matter bias'' is constrained to obey the consistency relation that it is unbiased with respect to itself
\be
\int dM \frac{dn}{dM }\frac{M}{\rho_m}b(M,z)=1;
\ee
the 1-halo term can also be written:
\be
P_{mm}^{1h}(k,z)=\int dM \frac{dn}{dM }\lb\frac{M}{\rho_m}u(k,M,z)\rb^2 .
\ee

The cross power between $\phi$ and the CIB is given by a Limber integration over the emissivity-matter cross-power spectrum $P_{mj}^\nu(k,z)$
\be
C_L^{\phi\nu}=\frac{2}{L^2}\int \frac{d\chi}{\chi^2}W_\kappa(\chi)  a(\chi) \bar j_\nu(z) P_{mj}^\nu\lb k=\frac{L}{\chi},z\rb,
\ee
where 
\begin{equation}
\bar j ^\nu(z)P_{mj}^\nu(k,z){}^{\rm 2-halo}=D_\nu(z)\lb\int dM \frac{dn}{dM }\frac{M}{\rho_m}b(M,z)u(k,M,z)\rb P_{\rm lin}(k,z)
\end{equation}
and
\begin{equation}
\bar j ^\nu(z)P_{mj}^\nu(k,z){}^{\rm 1-halo}=\int dM \frac{dn}{dM}\frac{M}{\rho_m}u(k,M,z)\frac{1}{4\pi}\lb L_{(1+z)\nu}^{\rm cen}+L_{(1+z)\nu}^{\rm sat}u(k,M,z)\rb.
\end{equation}
In all our lensing forecasts, we restrict the $L$ range to be $186\le L\le 1000$, where the lensing power spectrum is  in the linear regime, and so we take only the linearised 2-halo terms of the matter and matter cross emissivity spectra. It is worth noting, however, that the higher CIB frequencies --- which are sourced at lower redshift --- may have some contribution from the 1-halo term even at these scales; however we leave this issue to future modeling.

\section{A parametric $L-M$ relation}\label{sec:parametric_CIB}

To model the CIB power spectra one needs to specify the details of the halo model, and a luminosity-mass relation. For the halo model, we use the halo bias, halo mass function, and subhalo mass function of Tinker \cite{2010ApJ...724..878T, 2010ApJ...719...88T}.  We assume NFW halo profiles when calculating $u(k,m)$. The number of central galaxies $N^{\rm cen}(M,z)$ hosted by a halo of mass $M$ is modelled as
\be
N^{\rm cen}(M,z)=\begin{cases}0&M<M_{\rm min}\\1&M\ge M_{\rm min}
\end{cases}
\ee
where $M_{\rm min}$, the minimum halo mass to host a galaxy, is one of the parameters of the model; in the fiducial model, we use $M_{\rm min}=10^{10}M_\odot$.

The luminosity-mass relation we consider was introduced in \cite{2012MNRAS.421.2832S}; this parametric model has been fit to several data sets with various subsets of the parameters allowed to vary; for fits to SPIRE data see \cite{2013ApJ...772...77V} and for fits to \Planck data see \cite{2014A&A...571A..30P}.

The $L-M$ relation is parameterised by separating its dependence on mass and redshift and specifying
\be
L^{\rm gal}_{(1+z)\nu}=L_0 \Phi(z) \Sigma (M) \Theta ((1+z)\nu).\label{L_M}
\ee
$L_0$ is an overall normalisation factor which can be allowed to vary as a parameter in the model; $\Phi(z)$ determines the redshift evolution of the $L-M$ relation; $\Sigma(M)$ determines the mass dependence; and $\Theta$ is the spectral energy distribution (SED).  We will discuss these functions below. The fiducial values we quote for the CIB model parameters are the best-fit ones of \cite{2014A&A...571A..30P}\footnote{The value of $L_0$ is not listed in  \cite{2014A&A...571A..30P}  and so we choose a value that reproduces the amplitude of the CIB power spectra and intensities therein.}. These values are summarised in Sec. \ref{sec:parameter_values} and Table \ref{tab:fiducial_params}.

\subsubsection{Redshift evolution: $\Phi(z)$}

$\Phi(z)$ controls the redshift dependence of the normalisation of the $L-M$ relation and is parametrised by
\be
\Phi(z)=(1+z)^\delta.
\ee
We use $\delta=3.6$ for our fiducial model.

Various implementations \cite{2012MNRAS.421.2832S, 2013ApJ...772...77V} of this parametric model also consider another parameter $z_p$ at which the $L-M$ relation plateaus; in such a case
\be
\Phi(z)=\begin{cases}(1+z)^\delta&z<z_p\\
(1+z_p)^\delta&z\ge z_p\end{cases}.
\ee
This break is motivated by observational evidence of such a plateau in the $L-M$ relation, at $z\sim 2$. However, the model we consider does not include such a plateau.

\subsubsection{Mass dependence: $\Sigma(M)$}


$\Sigma (M)$ controls the dependence of luminosity on halo mass and is a log-normal function
\be
\Sigma (M) = \frac{M}{\sqrt{2\pi \sigma^2_{L/M}}}e^{-\lb\log_{10} M -\log_{10} M_{\rm eff}\rb^2/2\sigma ^2 _{L/M}}.
\ee
$\Sigma (M)$ is specified by two parameters: $M_{\rm eff}$, the peak of the specific IR emissivity ($L/M$); and $\sigma ^2_{L/M}$, which controls the range of halo masses that produce the emissivity.  In our fiducial model, we use $M_{\rm eff}=10^{12.6}M_\odot$ and $\sigma ^2_{L/M}=0.5$. Note that $\sigma ^2_{L/M}$ was not varied in the analysis of \cite{2014A&A...571A..30P} and is considered fixed in our forecasts.

\subsubsection{IR SEDs: $\Theta(\nu,z)$}\label{sec:IR_SED}

Finally, the SED $\Theta$ is a modified black body with a power-law tail at high frequencies
\be
\Theta \propto \begin{cases} \nu^\beta B_\nu(T_d(z))&\nu<\nu_0\\
\nu^{-\gamma} &\nu\ge\nu_0\end{cases}\label{SED}
\ee
where $B_\nu(T)$ is the Planck function at temperature $T$ and $T_d(z)$ is the dust temperature at redshift $z$. $\nu_0$ is the ($z$-dependent) frequency satisfying the continuous derivative relation
\be
\frac{d\ln \Theta(\nu,z)}{d\ln\nu}\bigg{|}_{\nu=\nu_0}=-\gamma.
\ee
We take $\beta =1.75$ and $\gamma = 1.7$ in our fiducial model.

The dust temperature is parametrised as 
\be
T_d=T_0\lb1+z\rb^\alpha.
\ee
In the fiducial model, $T_0=24.4 \,\mathrm{K}$ and $\alpha=0.36$.
 
Thus there are four parameters that control the SED: the gray-body emissivity factor $\beta$, the high-frequency power-law exponent $\gamma$, the dust temperature today $T_0$, and $\alpha$, which controls the redshift evolution of the temperature. Note that SEDs are normalised such that $\Theta(\nu_0)=1$, i.e. such that
\be
\Theta = \begin{cases} \lb\frac{\nu}{\nu_0}\rb^\beta \frac{B_\nu(T_d(z))}{B_{\nu_0}(T_d(z))}&\nu<\nu_0\\
\lb\frac{\nu}{\nu_0}\rb^{-\gamma} &\nu\ge\nu_0.\end{cases}\label{SED_parametric}
\ee
Plots of the SEDs at various fixed redshifts are shown in Figure \ref{fig:redshift_distribution}.

\subsubsection{Values of the parameters}\label{sec:parameter_values}

We consider the model of \cite{2014A&A...571A..30P}, which was fit to the CIB power spectra at \{217, 353, 545, 857, 3000\} GHz. In this model, the parameters $\delta$, $\beta$, $T_0$, $\gamma$, $\alpha$ ,$\log_{10}{M_{\rm min}}$, $\log_{10}{M_{\rm eff}}$, and $L_0$ were varied; $\sigma_{L/M}^2$ was fixed at $0.5$. There is no plateau in the $L-M$ relation: $\Phi(z)$ behaves as $(1+z)^\delta$ at all redshifts. The parameters are summarised and their values are given in Table \ref{tab:fiducial_params}. The values of the shot noises in the power spectra were also allowed vary as parameters, and marginalised over; their best-fit values are given in Table \ref{tab:shot_noises}. With the SED normalised as in Equation \eqref{SED_parametric}, the fiducial value we use for $L_0$ is  $L_0 = 6.4\times10^{-8}\text{Jy MPc}^2 /M_\odot / \text{Hz} =1.49\times10^{-15} L_\odot/M_\odot/\text{Hz}$.

\begin{table*}[t]
\begin{tabular}{|c|l|c|}
\hline
Parameter & Parameter description & Value\\
\hline\hline
$\alpha$&Redshift evolution of dust temperature  & $0.36\pm0.05$\\\hline
$T_0$ & Dust temperature at $z=0$ & $24.4\pm1.9$ K\\\hline
$\beta$ & Emissivity index of SED & $1.75\pm0.06$\\\hline
$\gamma$ & Power law index of SED at high frequency & $1.7\pm0.2$\\\hline
$\delta$ & Redshift evolution of $L-M$ normalisation & $3.6\pm0.2$\\\hline
$\log_{10}M_{\rm eff}/M_\odot$&Most efficient halo mass & $12.6\pm0.1$\\\hline
$\log_{10}M_{\rm min}/M_\odot$&Minimum halo mass to host a galaxy & unconstrained\\\hline
$L_0$&Normalisation of $L-M$ relation&$6.4\times10^{-8}\text{Jy MPc}^2 /M_\odot / \text{Hz} $\\\hline
$\sigma^2_{L/M}$&Size of of halo masses sourcing CIB emission & 0.5 (not varied)\\
\hline
\end{tabular}
\caption{Best-fit parameters of \cite{2014A&A...571A..30P}}\label{tab:fiducial_params}
\end{table*}
\begin{table}[h!]
\begin{tabular}{|c||c|c|c|c|c|}
\hline
$\nu,\nu^\prime$&217 & 353 & 545 & 857 & 3000\\\hline\hline
217 &  21 & 54 & 121 & 181 & 95 \\\hline
353 &  & 262 & 626 & 953 & 411\\\hline
545 &  & &1690 & 2702 & 1449\\\hline
857 &  & & & 5364& 4158\\\hline
3000 &  & & & &9585\\\hline
\end{tabular}
\caption{Shot noise values of  \cite{2014A&A...571A..30P}, in $\mathrm{Jy^2/sr}.$ The frequencies are in GHz and are the frequencies for which the CIB power spectra were measured and used to fit the model.}\label{tab:shot_noises}
\end{table}

\begin{figure}[h!]
\includegraphics[width=0.6\textwidth]{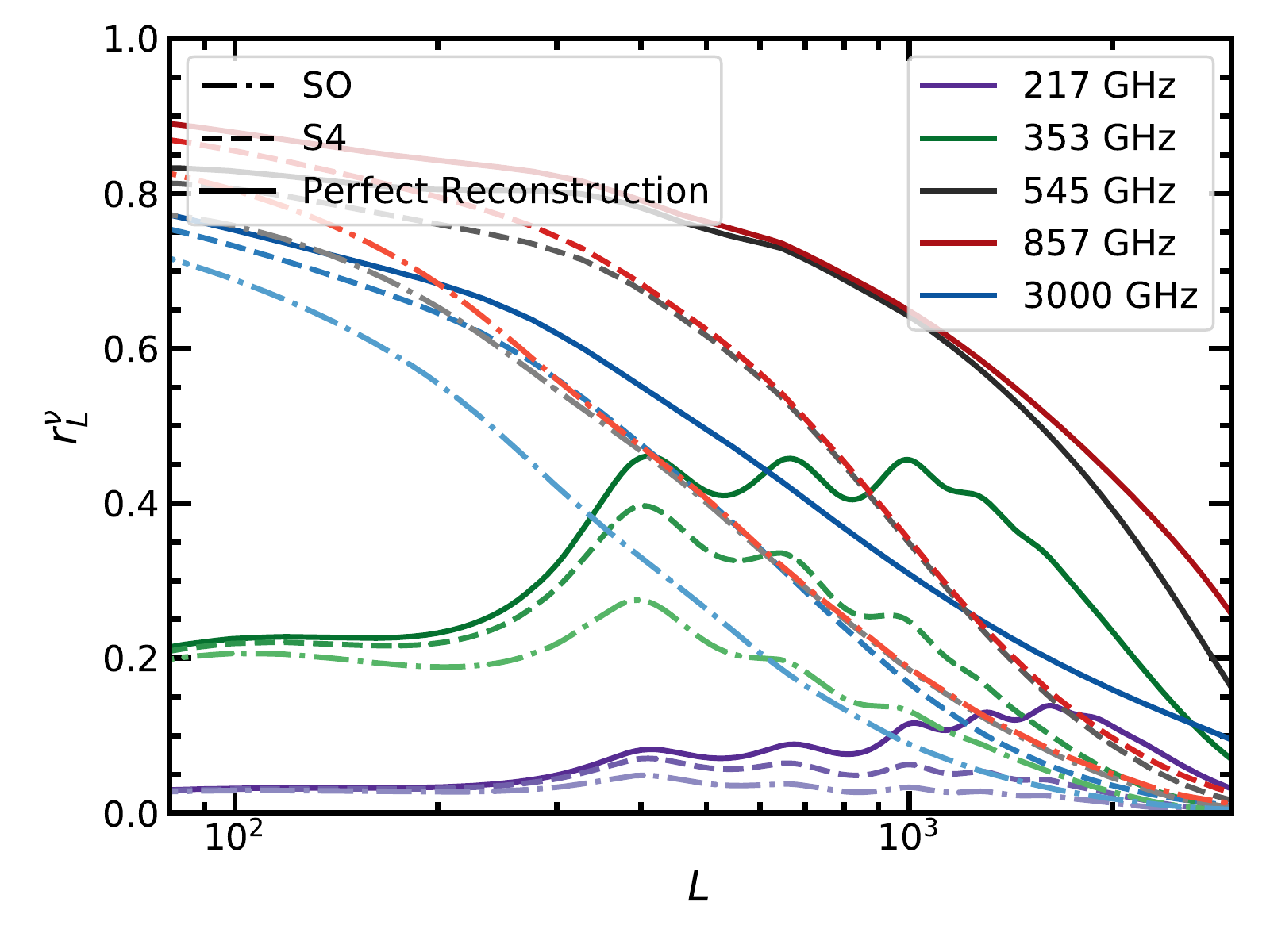}
\caption{The correlation coefficients between the CIB maps from \Planck+IRIS with CMB lensing maps from various experiments. The calculation includes instrumental and foreground noise in the CIB maps. The perfect reconstruction case corresponds to no noise on the CMB lensing reconstruction.}\label{fig:corrcoeff}
\end{figure}
\section{Fisher Forecasts}\label{sec:fisher_forecasts}

We perform various Fisher forecasts to investigate whether inclusion of CIB/CMB lensing data has power to improve constraints on CIB models. We consider two experimental configurations for the CIB: one corresponding to the \Planck+IRIS experiments, at \{217, 353, 545, 857, 3000\} GHz, and one corresponding to the upcoming CCAT-prime \cite{2020JLTP..199.1089C} survey, which will measure the CIB on small angular scales at \{220,  280,  350,  410,  850\} GHz.  

We are considering improvements in only the parameters of the CIB model; the CMB lensing power spectrum is not dependent on any of the parameters we are including in our forecast\footnote{In particular, we assume cosmological parameters are known to much better precision than the CIB model parameters considered here.}. However, the CMB lensing power spectrum and the CIB power spectra are correlated;  see Figure \ref{fig:corrcoeff} for plots of the correlation coefficients $r_L^\nu\equiv \frac{C_L^{\nu\phi}}{\sqrt{C_L^{\nu\nu}C_L^{\phi\phi}}}$. Due to the high correlation coefficient, measuring these two fields \ti{on the same patch of sky} can yield improvements in a model describing one field even if the other is not dependent on this model through the cancellation of sample-variance shared by the two fields \cite{Seljak:2008xr}. Additionally, knowledge of the redshift distribution of the CIB is contained in the correlation or lack thereof \cite{2013arXiv1303.4722M} with the CMB lensing matter distribution, whose redshift dependence is precisely known. Due to this, we expect the inclusion of CMB lensing in the CIB analysis to yield improvements in the CIB parameters.

\subsection{Fisher matrix formalism}\label{sec:fisher}
\begin{figure*}[t]
\includegraphics[width=0.5\textwidth]{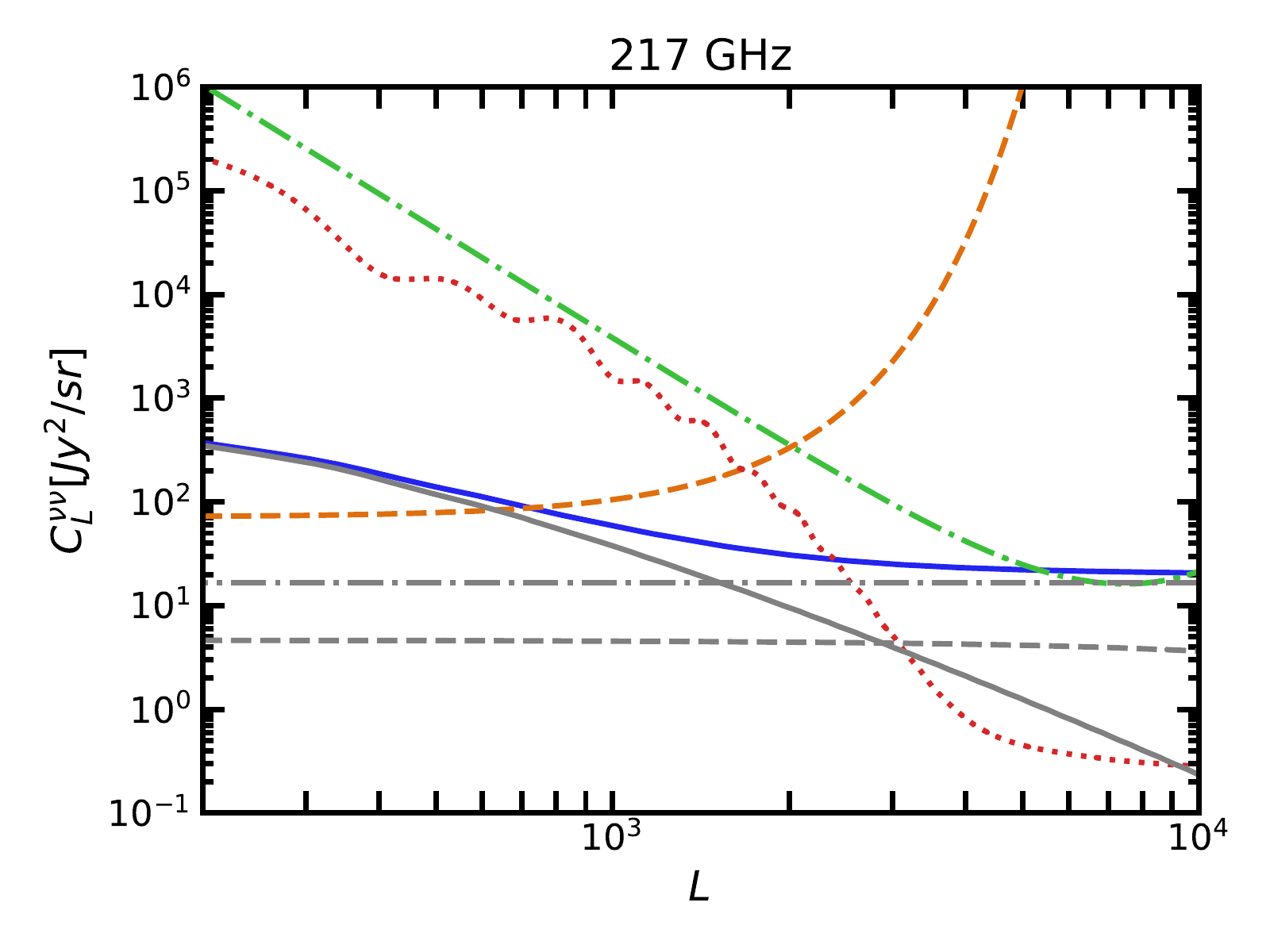}\includegraphics[width=0.5\textwidth]{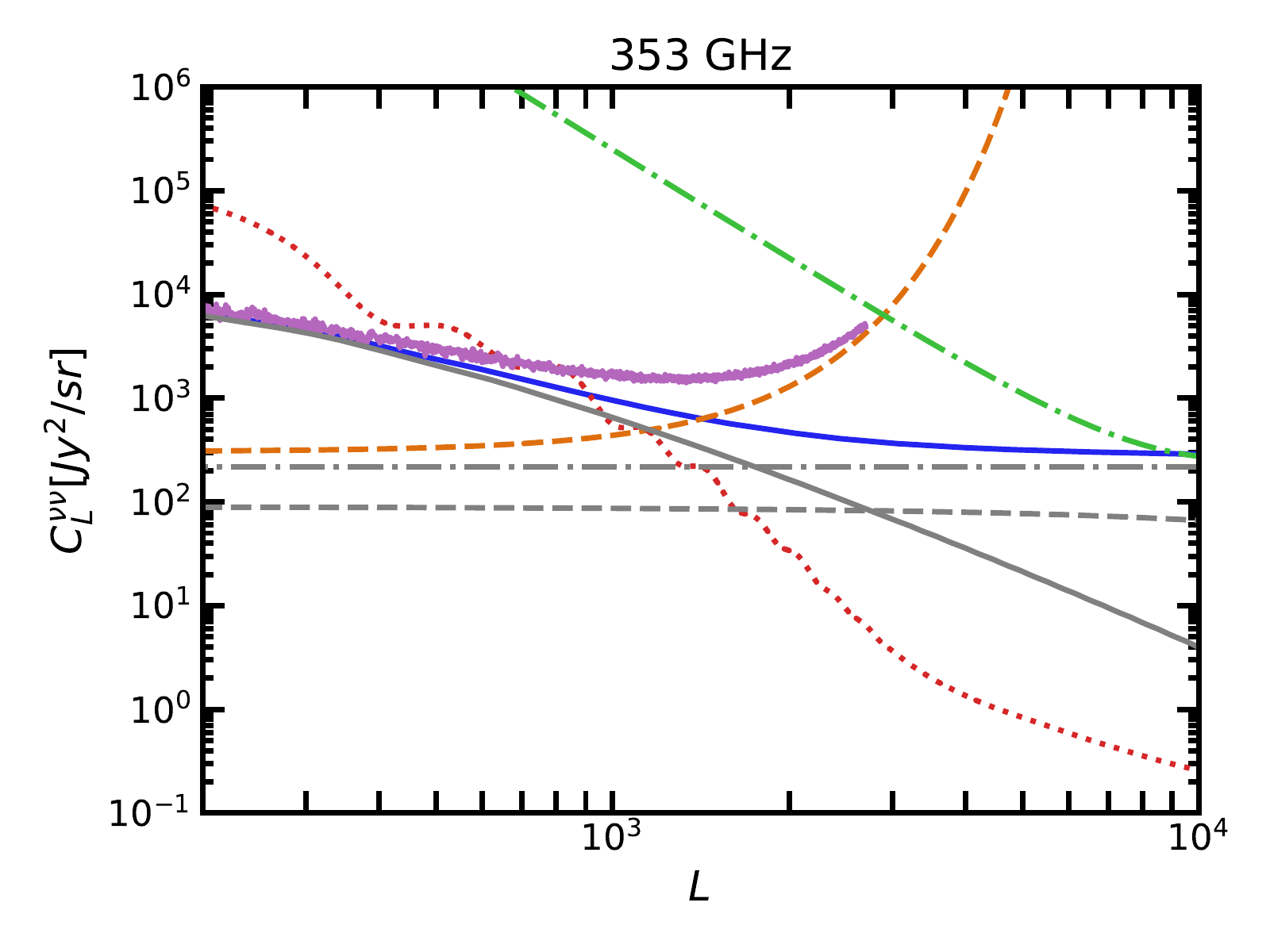}\\
\includegraphics[width=0.5\textwidth]{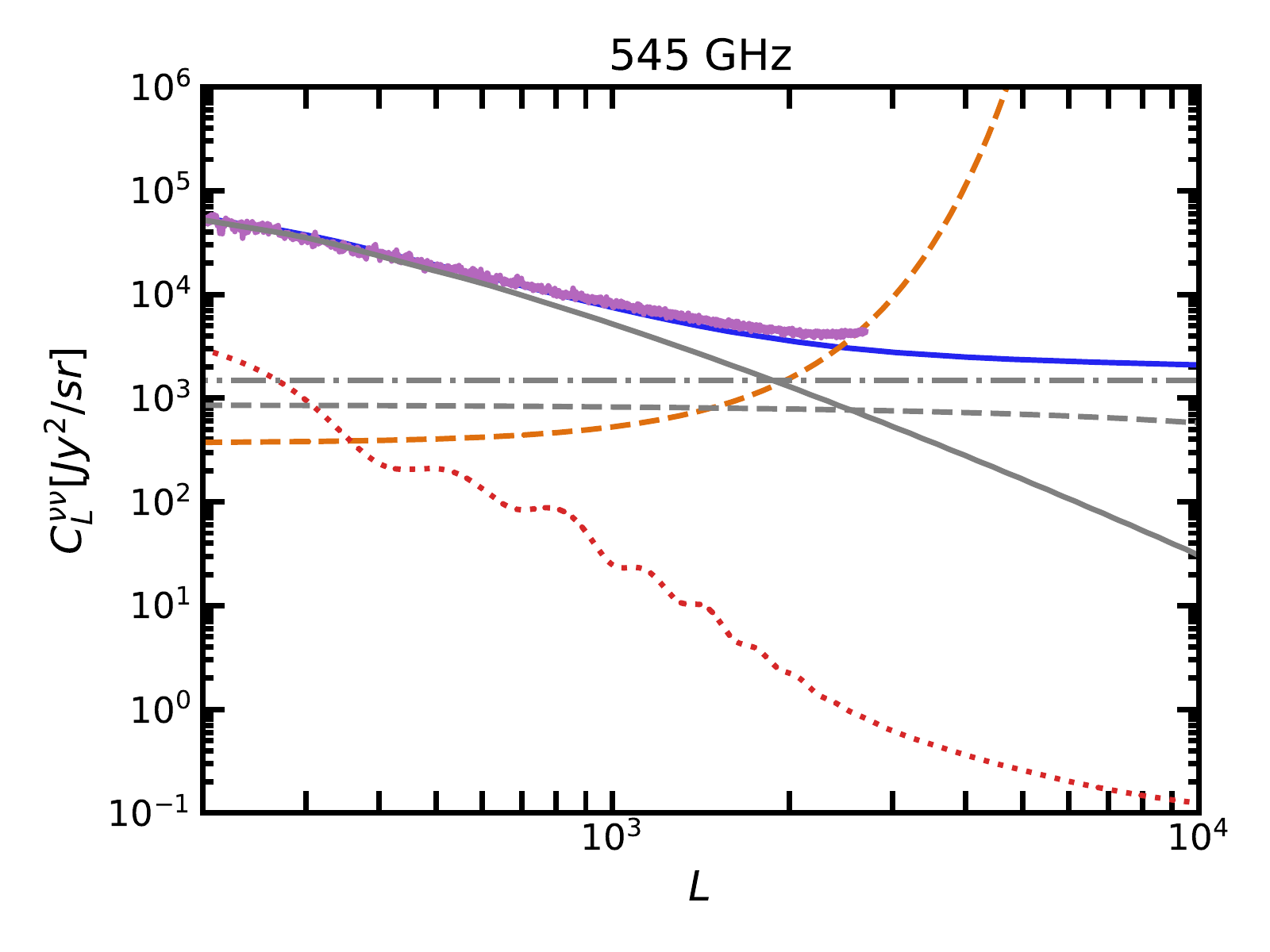}\includegraphics[width=0.5\textwidth]{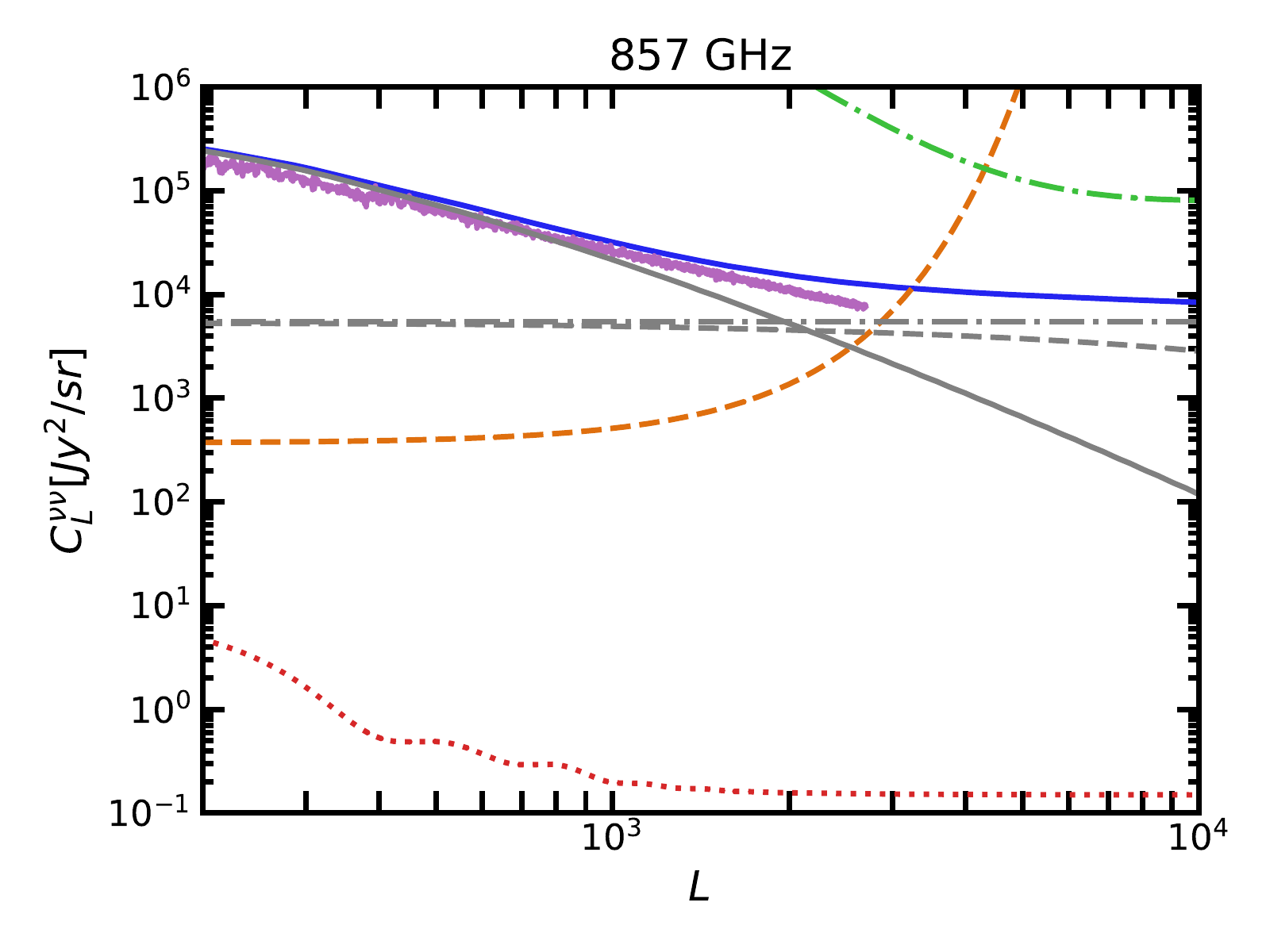}\\
\includegraphics[width=0.5\textwidth]{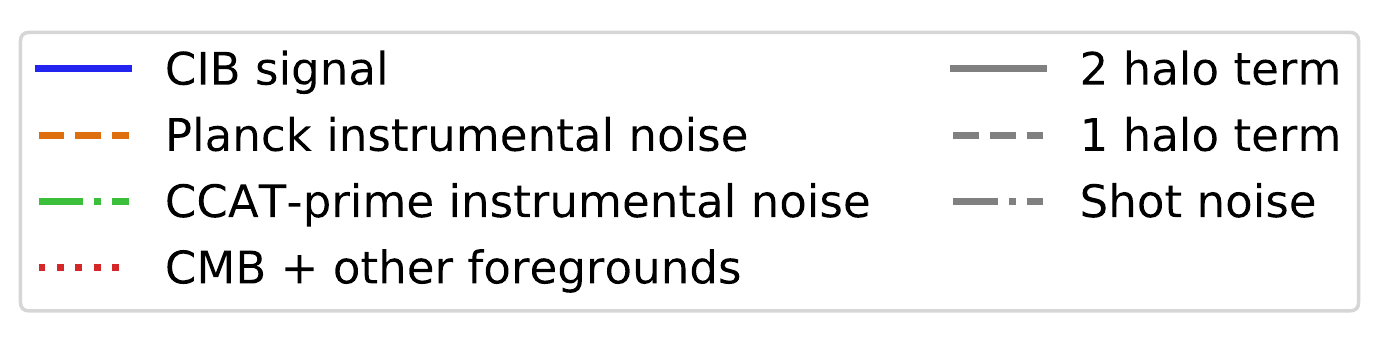}
\caption{The predicted CIB signal at \Planck frequencies is shown here in solid blue along with beam-deconvolved instrumental noise (orange dashed) and total foreground power (red dotted). Also shown in purple are the power spectra of the CIB maps of \cite{Lenz:2019ugy} (for 353, 545 and 857 GHz), corrected for the beam and partial sky coverage. When available at a nearby frequency, the beam-deconvolved CCAT-prime instrumental noise (one of many possible configurations) is also shown. This figure shows that our signal and noise power spectra account fairly well for the observed CIB power in \cite{Lenz:2019ugy}.}\label{fig:planck_signal_and_noise}
\end{figure*}

\begin{figure*}[t]
\includegraphics[width=0.48\textwidth]{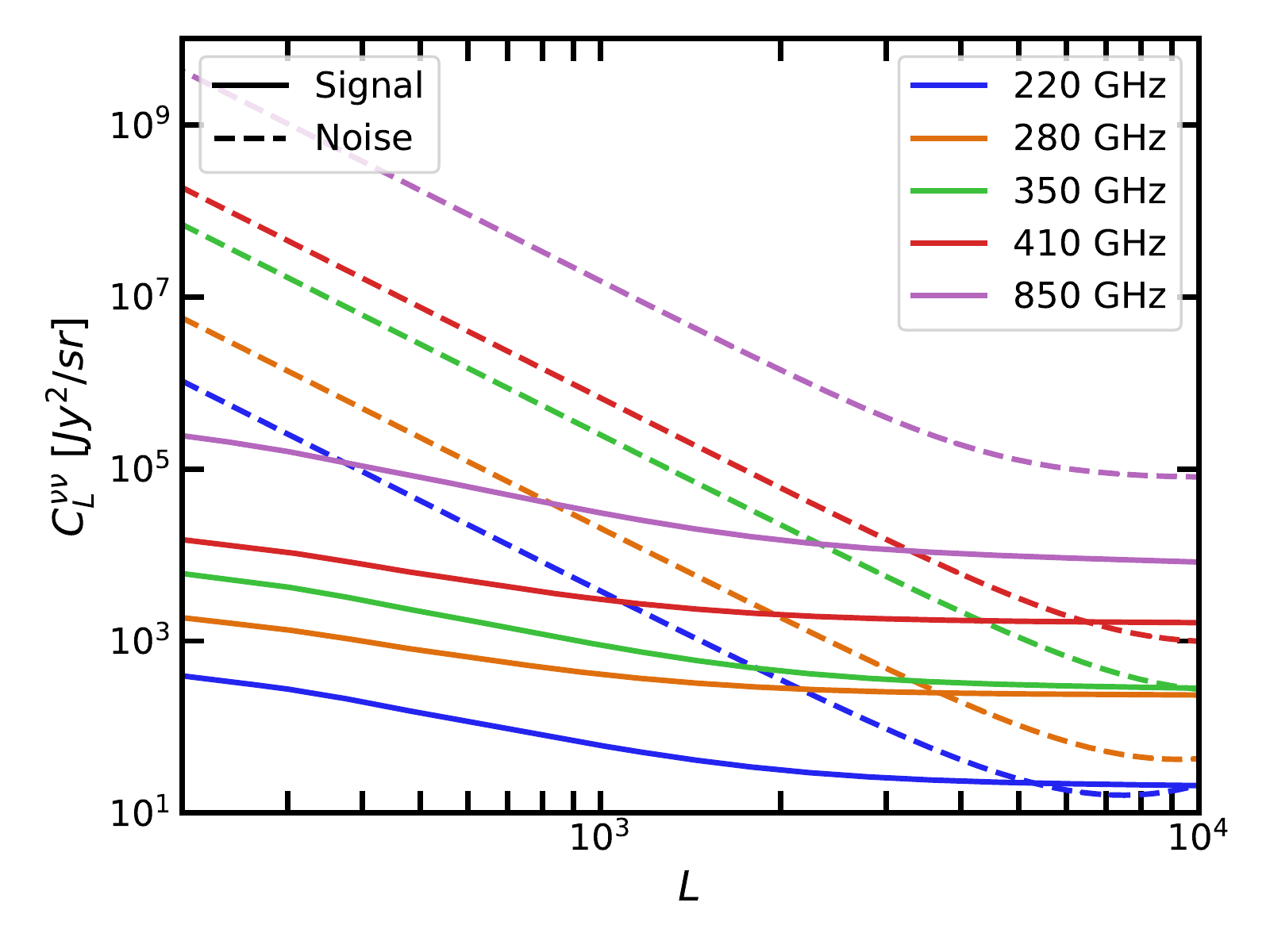} 
\includegraphics[width=0.48\textwidth]{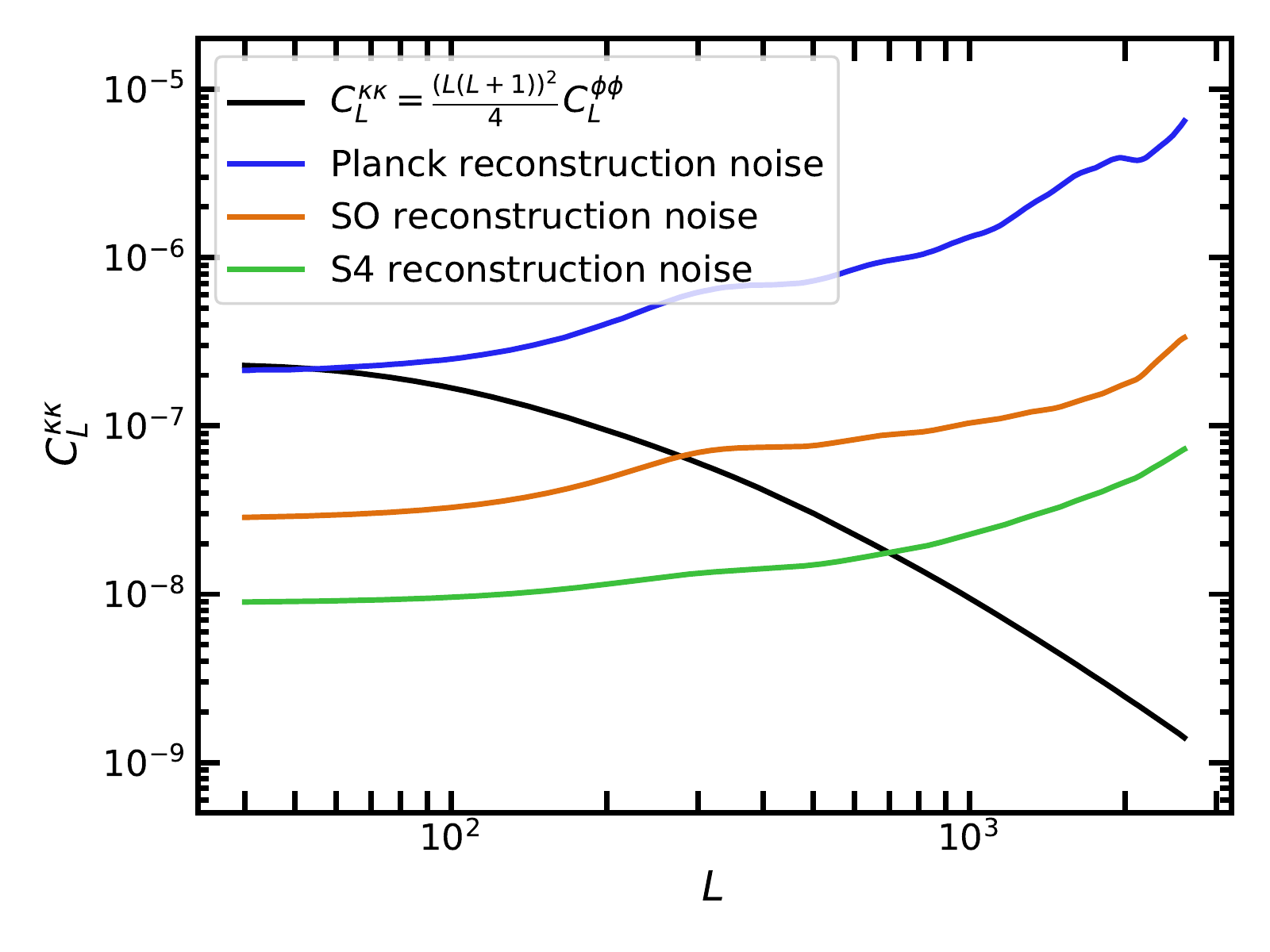}
\caption{{\it Left:} The CIB signal at CCAT-prime frequencies (solid) shown against the beam-deconvolved instrument noise for one of many possible configurations of CCAT-prime, corresponding to the values in Table \ref{tab:noise_planck}. {\it Right:} The CMB lensing convergence power spectrum shown against various reconstruction noise levels from the \Planck satellite, a Simons Observatory-like configuration and a CMB-S4-like configuration (note that we plot the lensing convergence power spectrum which is related to the lensing potential power spectrum through $C_L^{\kappa\kappa}=\frac{\lb L(L+1)\rb^2}{4}C_L^{\phi\phi})$.}\label{fig:ccatp_signal_and_noise}
\end{figure*}

\begin{figure*}[t]
\includegraphics[width=0.5\textwidth]{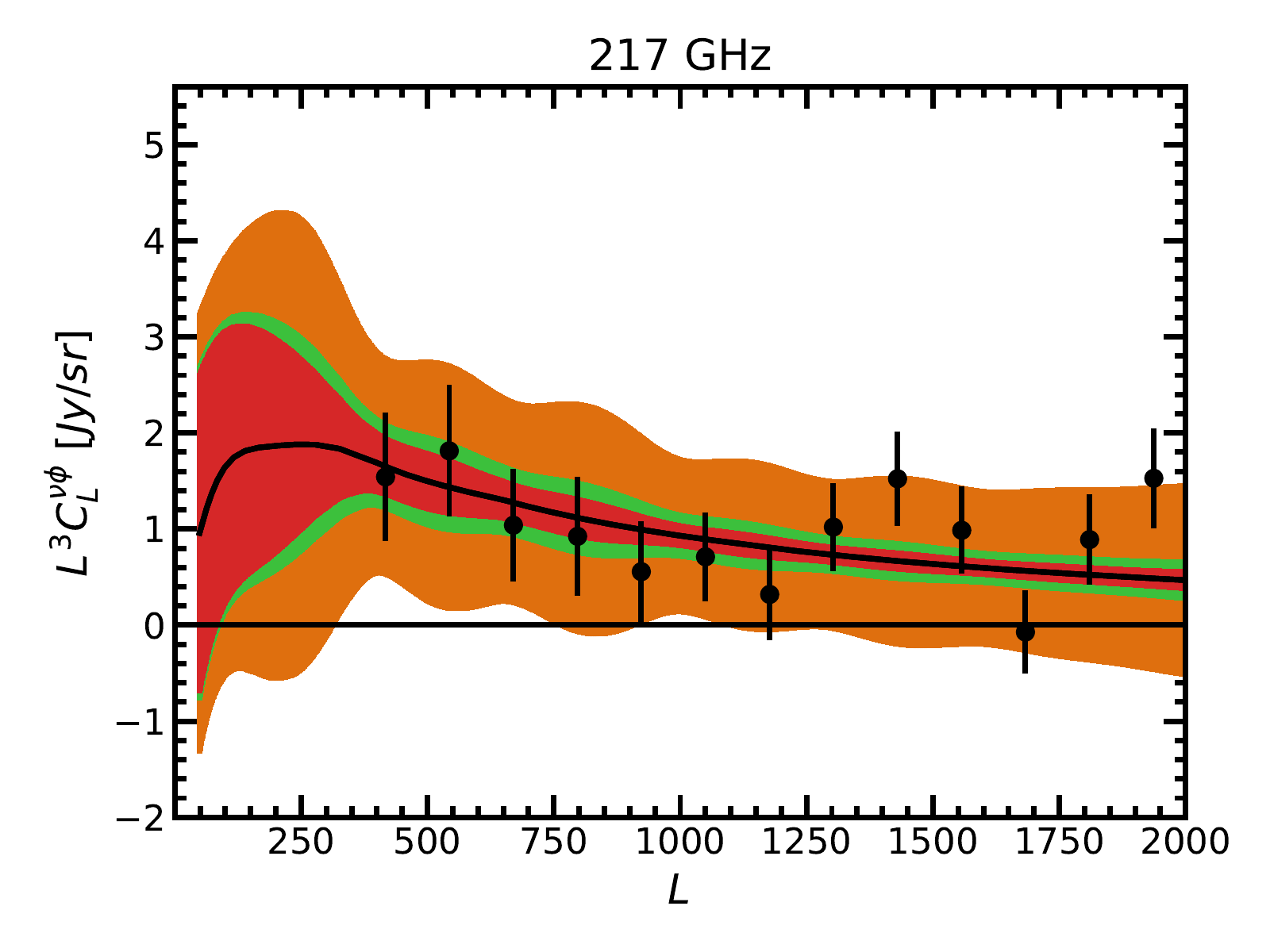}\includegraphics[width=0.5\textwidth]{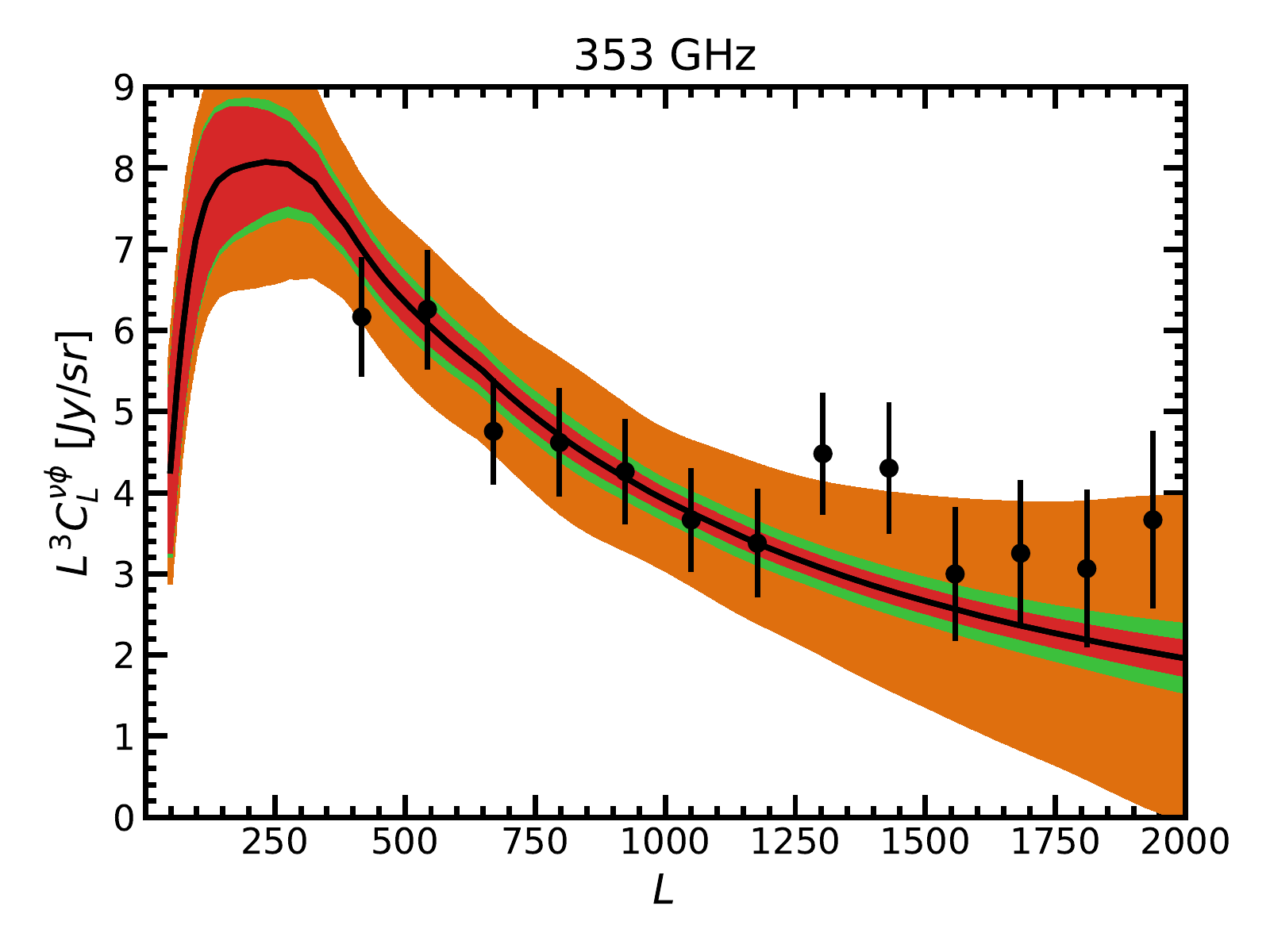}\\
\includegraphics[width=0.5\textwidth]{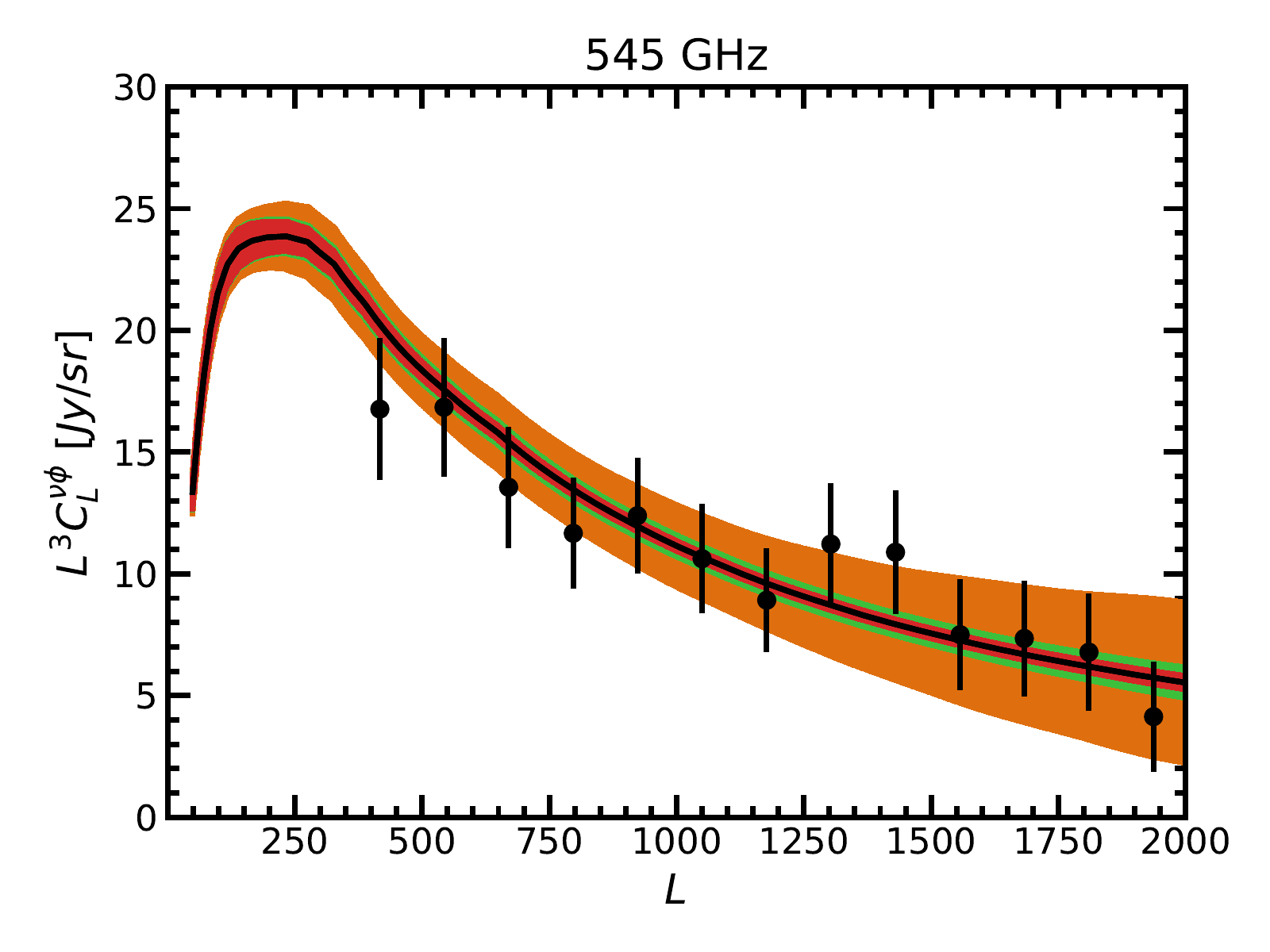}\includegraphics[width=0.5\textwidth]{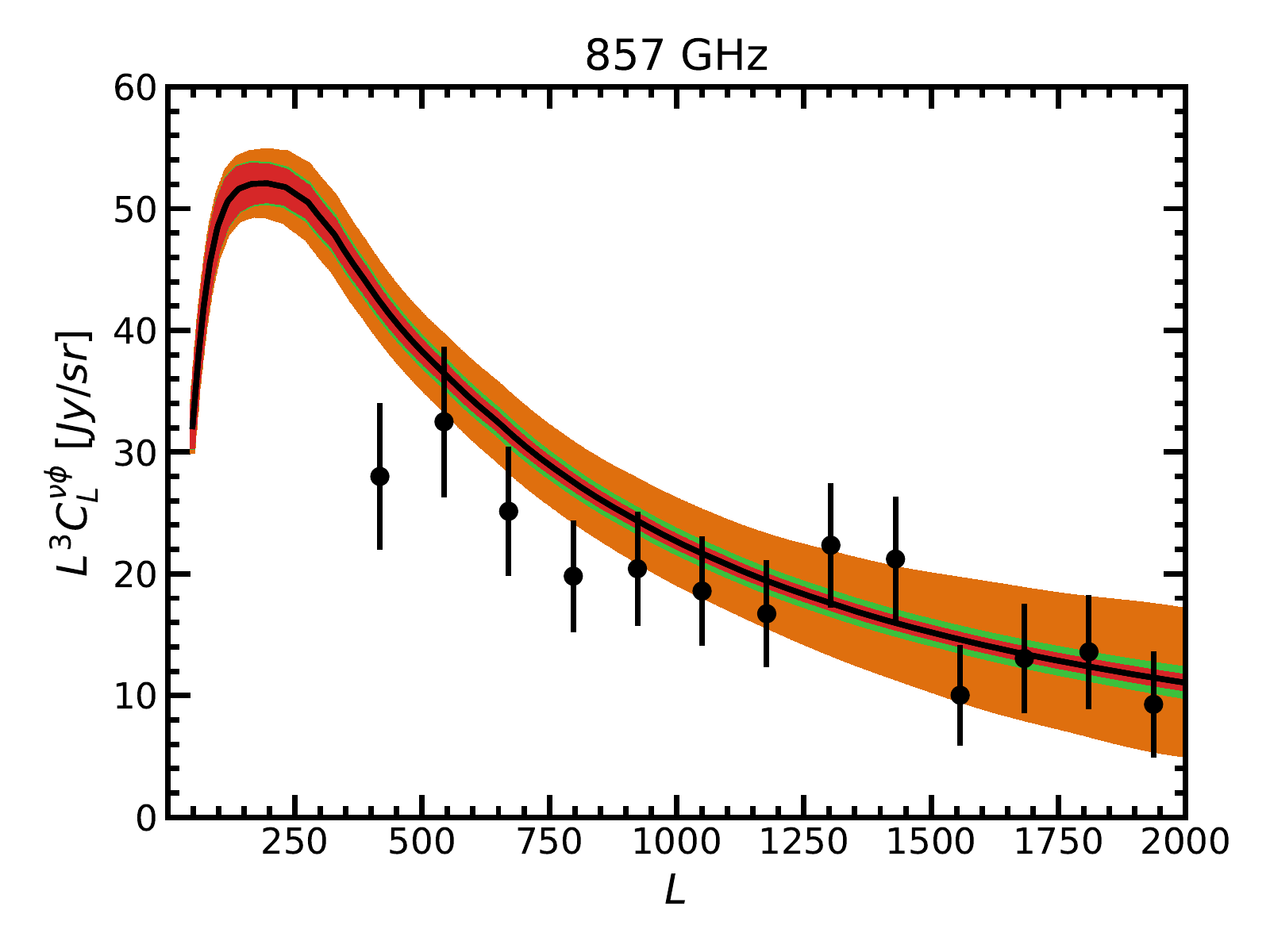}
\caption{The cross-power spectrum between CMB lensing and the CIB measured by \Planck at various frequencies. The model prediction for $L^3C_L^{\phi\nu}$ is plotted in black. The data points show the measurements using \Planck lensing from \cite{Ade:2013aro}.  The predicted 1-$\sigma$ uncertainty lensing reconstruction noise and our CIB model is shown in orange for \Planck, in bins of width $\Delta L= 126$; in red for a Simons Observatory-like lensing reconstruction, and in green for a CMB-S4-like lensing reconstruction. Note that the error bars on the \Planck data points are smaller than predicted at low frequency, as our analysis (conservatively) does not assume that the CMB is cleaned using lower frequency data.}\label{fig:cross_errorbars}
\end{figure*}

\begin{figure*}[t]
\hspace{1em}
\includegraphics[width=0.99\textwidth]{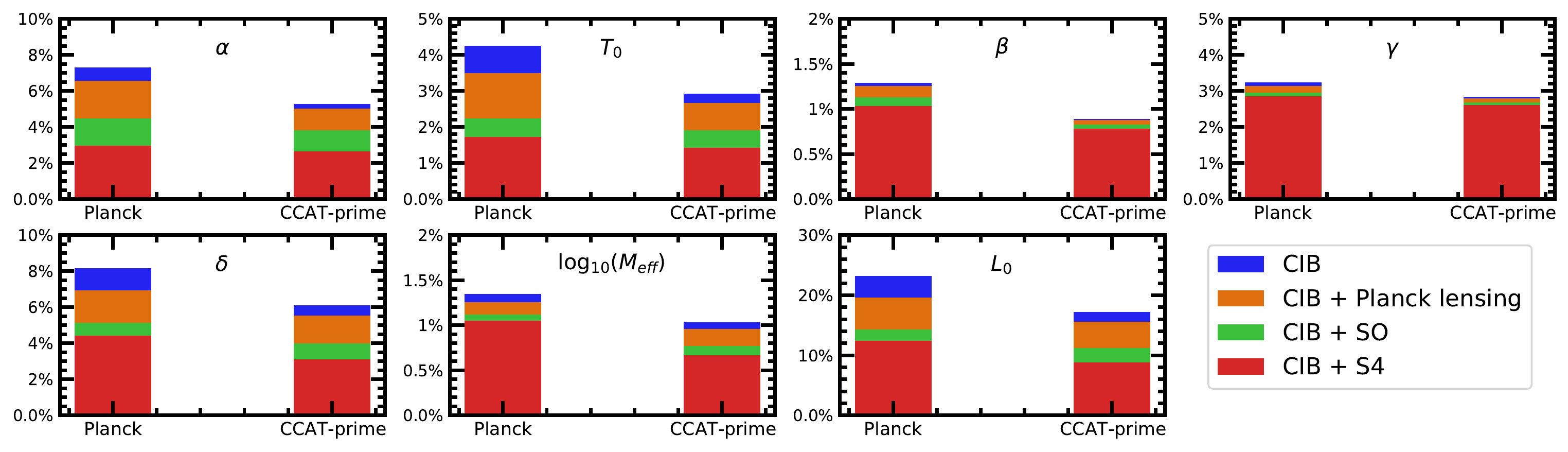}
\caption{Percentage constraints on various CIB halo model parameters and their improvement with the incorporation of CMB lensing, for the CIB as measured by \Planck and by CCAT-prime. We show improvements when including lensing reconstruction from \Planck itself or from a future Simons Observatory-like or CMB-S4-like survey configuration. }\label{fig:bargraph}
\end{figure*}

\begin{figure*}[t]
\includegraphics[width=0.99\textwidth]{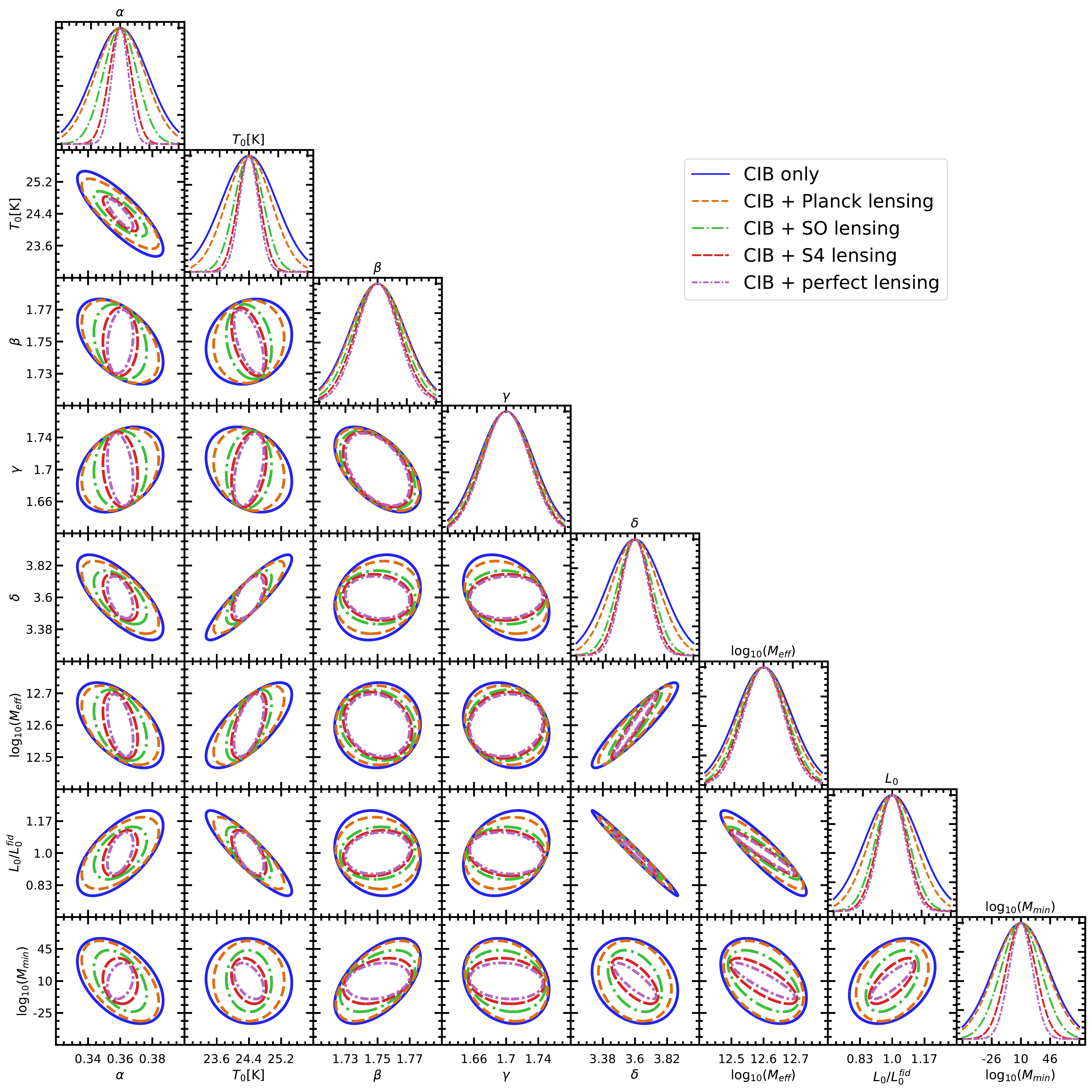}
\caption{Forecast 1-$\sigma$ confidence ellipses for various parameters of the CIB model, with and without CMB lensing information:  in blue solid, we show constraints when only including \Planck CIB measurements. In orange dashed, we show constraints when including \Planck lensing reconstruction in addition. In green (dot-dashed) and red (dashed), the constraints when adding a Simons Observatory-like and CMB-S4-like lensing reconstruction are shown respectively. In purple dot-dashed, we show constraints when adding a noiseless CMB lensing reconstruction.}\label{fig:triangleplanck}
\end{figure*}

\begin{figure*}[t]
\includegraphics[width=0.5\textwidth]{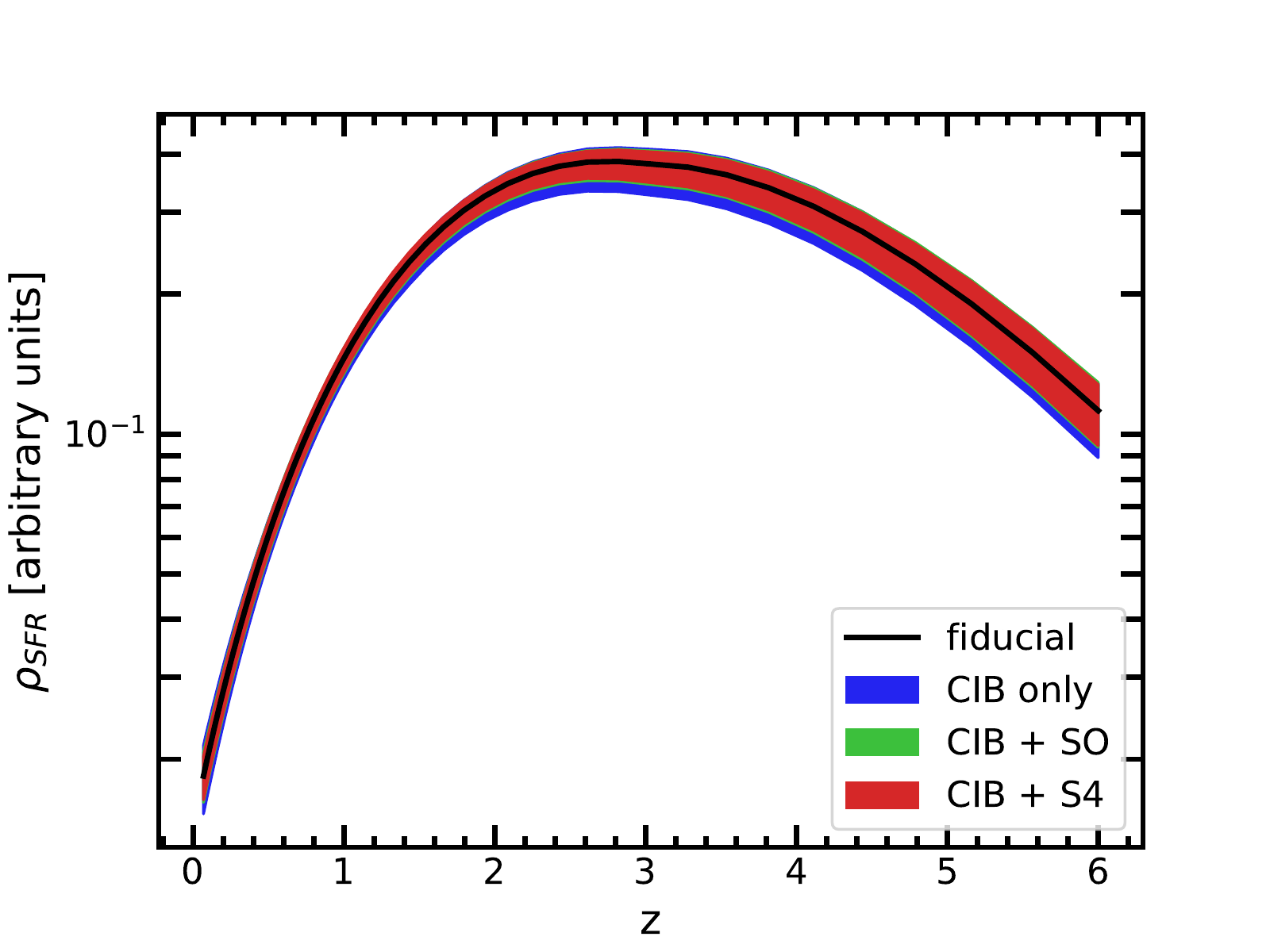}\includegraphics[width=0.5\textwidth]{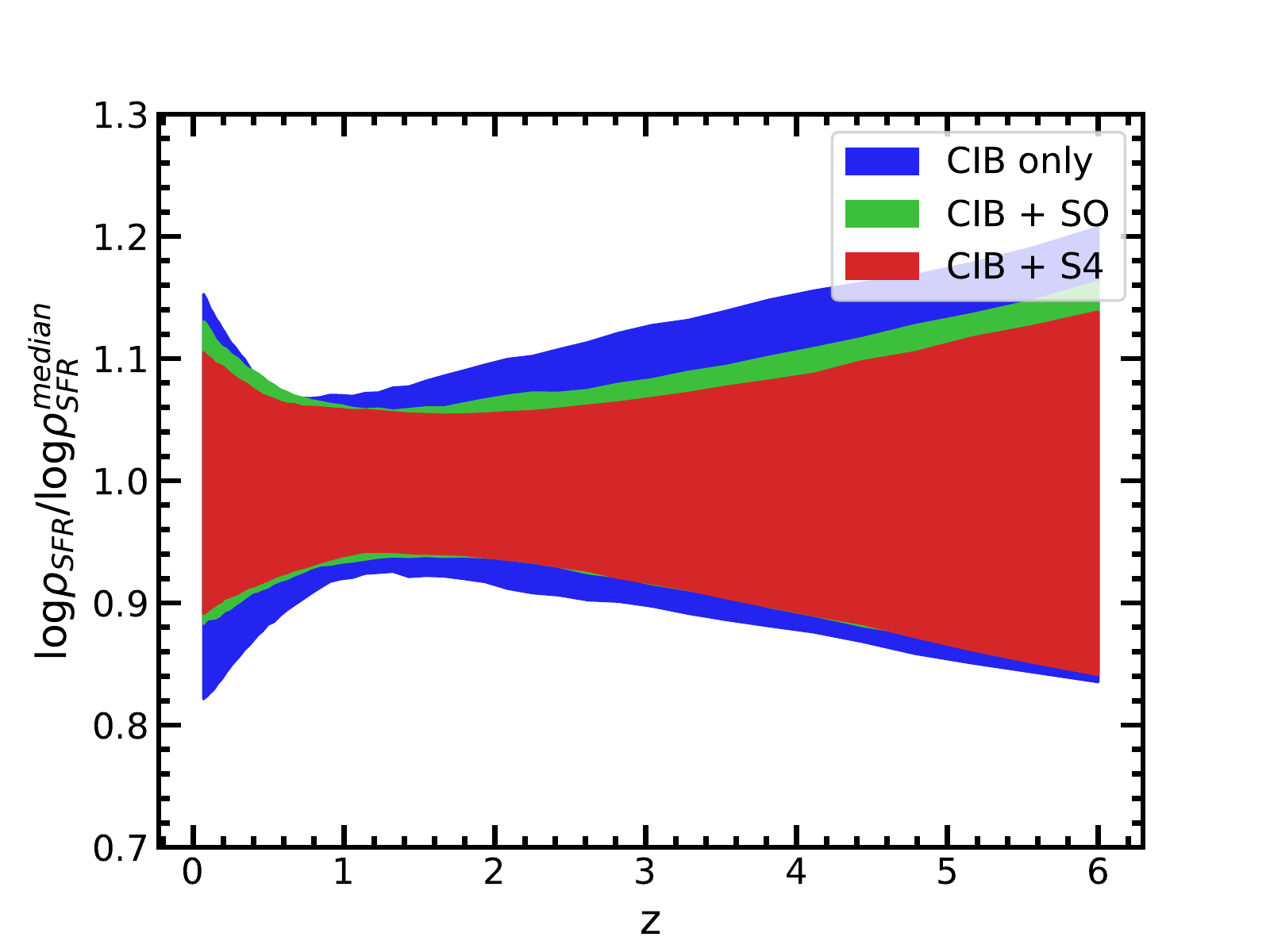}

\caption{Star formation rate constraints for \Planck-like CIB data with various lensing scenarios. On the left $\rho_{\rm SFR}$ is plotted in arbitrary units on a log scale, and the fiducial value is also shown. On the right we show a linear scale, and we divide by the median value of each set of realisations to make the improvements more visible. Note that we only consider \Planck-like CIB data, specifically the no-lensing scenario (blue), the SO-lensing scenario (green), and the S4-like lensing scenario (red). }\label{fig:sfr}
\end{figure*}

We consider at each $L$ an $(N+1)\times (N+1)$ covariance matrix , where $N$ is the number of frequency channels at which the CIB is measured:
\be
C_L=\left(
\begin{array}{c c}
C_L^{\nu\nu^\prime}&C_L^{\nu\phi}\\
C_L^{\nu\phi}&C_L^{\phi\phi}
\end{array}
\right).
\ee
$C_L^{\nu\nu^\prime}$ is an $N\times N$ covariance matrix of the auto- and cross-power spectra of the CIB, $C_L^{\nu\phi}$ is an $N$-dimensional vector of the cross power spectra between the CIB and CMB lensing, and $C_L^{\phi\phi}$ is the CMB lensing power spectrum. 

We consider a vector of parameters
\begin{align}
\Pi^i = \big{(}\alpha,T_0,\beta,\gamma,&\delta, \log_{10}M_{\rm eff}/M_\odot,\log_{10}M_{\rm min}/M_\odot,L_0,S^{\nu\times\nu^\prime}\big{)}
\end{align}
where $S^{\nu\times\nu^\prime}$ denotes the $\frac{N(N+1)}{2}$ shot noise parameters. 
The Fisher matrix for the parameters is defined as
\be
F_{ij}=\sum_{L}\frac{\lb2L+1\rb}{2}f_{\rm sky}\Tr\lsb C_L^{-1} \frac{\partial C_L}{\partial \Pi^i} C_L^{-1}\frac{\partial C_L}{\partial \Pi^j}\rsb\label{fishermatrix}
\ee
where $f_{\rm sky}$ is the sky fraction covered by the experiment. $C_L^{-1}$ includes both signal and noise.
Within this setup, the fully marginalised forecast $1\sigma$ error on a parameter $i$ is given by $\sqrt{\lb F^{-1}\rb_{ii}}$. 

At times, we refer to forecasts where we remove the CIB auto  power spectra from our forecasts, and consider only the cross power spectra $C_L^{\phi\nu}$ and the lensing auto-power $C_L^{\phi\phi}$. We do this by instead employing the bandpower Fisher formalism, where the Fisher matrix is computed from the covariance of the power spectra; in this case the data is considered to be the power spectra (as opposed to the fields themselves) and we consider the data vector 
\be
C_L=\left(C_L^{\nu,\nu^\prime},C_L^{\nu,\phi},C_L^{\phi\phi}\right)\label{Cl_datavector}
\ee
with covariance matrix 
\begin{widetext}
\begin{align}
\mathbb{C}\left(\hat{C}_{L}^{\alpha \beta}, \hat{C}_{L}^{\gamma \delta}\right)= \frac{1}{(2 L+1) f_{\mathrm{sky}}}\bigg{[}\left(C_{L}^{\alpha \gamma}
+ N_{L}^{\alpha \gamma}\right)\left(C_{L}^{\beta \delta}+N_{L}^{\beta \delta}\right)
+\left(C_{L}^{\alpha \delta}+N_{L}^{\alpha \delta}\right)\left(C_{L}^{\beta \gamma}+N_{L}^{\beta \gamma}\right)\bigg{]};\label{cov_bandpower}
\end{align}
\end{widetext}
$F_{ij}$ is now given by
\be
F_{ij}=\sum_L \frac{\partial C_L^T}{\partial \Pi^i }\mathbb{C}_L^{-1} \frac{\partial C_L}{\partial \Pi^j }
\ee
(where $C_L^T$ denotes the transpose of $C_L$).
This Fisher formalism is equivalent to the one that results in Equation \eqref{fishermatrix}; however, it allows us to explicitly remove power spectra from the analysis by taking only the entries we are interested in in \eqref{Cl_datavector}, something is not possible when we are computing Equation  \eqref{fishermatrix}.
\subsection{The CIB power spectra and CMB lensing: signal and noise}\label{sec:instrumental_noise}

\subsubsection{CIB noise specifications}
There is instrumental noise in the $\nu=\nu^\prime$ power  spectra which we include as
\be
N_L=N(L) e^{\frac{L(L+1)\Theta_{\rm FWHM}^2}{8\ln2}}
\ee
where $N(L)$ is the value of the  noise and $\Theta_{\rm FWHM}$ is the beam Full Width at Half Maximum in radians. For \Planck and IRIS we use only white noise $N(L)=N_{\rm white}$. The values of $N_{\rm white}$ and $\Theta_{\rm FWHM}$ are given in Table \ref{tab:noise_planck}. For CCAT-prime we consider  both large-scale frequency dependent ``red'' noise and white noise corresponding to the specifications given in \cite{2020JLTP..199.1089C}:
\be
N(L)=N_{\rm red}\lb\frac{L}{L_{\rm knee}}\rb^{\alpha_{\rm knee}}+N_{\rm white},
\ee
where $L_{\rm knee}$ is 1000 and $\alpha_{\rm knee}$ is 3.5. We emphasize that just as with Simons Observatory and CMB-S4 configurations considered below, the noise specifications correspond to one of many possible configurations that CCAT-prime could observe in. The values of $N_{\rm red}$, $N_{\rm white}$, and $\Theta_{\rm FWHM}$ are given in Table \ref{tab:noise_planck} \footnote{At \Planck frequencies, we change between $\mathrm{\mu K}$ and $\mathrm {Jy}$ using the conversion factors in \cite{2014A&A...571A...9P}. See Appendix \ref{App:units} for more details.} (note that we consider the noise levels corresponding to the configuration in which CCAT-prime observes 15,000 sq. deg., but we only use 2240 sq. deg. of these). We show in Figure \ref{fig:planck_signal_and_noise} the signal and noise at the frequencies measured by \Planck. We show in Figure \ref{fig:ccatp_signal_and_noise} a summary of the signal and noise at all of the CCAT-prime frequencies.

\begin{table*}[t]
\begin{center}
\begin{tabular}{|c||c||c|c|}
\hline
    \multirow{2}{*}{Frequency}& \multirow{2}{*}{Beam (arcmin)}&\multicolumn{2}{c|}{Noise}\\\cline{3-4}&&$\mathrm{Jy}^2/\mathrm{sr}$&$\mu \text{K-arcmin}$\\\hline\hline
    217 GHz &5.01&72&60.12\\\hline
353 GHz &4.86&305&208.98\\\hline
545 GHz &4.84&369&1137\\\hline
857 GHz &4.63&369&29075\\\hline
3000 GHz & 4.3  &305&$6.7\times10^{18}$\\
\hline
\end{tabular}\\
\begin{tabular}{|c||c||c|c||c|c|}
\hline
  \multirow{2}{*}{Frequency}&\multirow{2}{*}{Beam (arcmin)}&\multicolumn{2}{c||}{White noise}&\multicolumn{2}{c|}{`Red' noise}\\\cline{3-6}&&$\mathrm{Jy}^2/\mathrm{sr}$&$\mu \text{K-arcmin}$&$\mathrm{Jy}^2/\mathrm{sr}$&$\mu \text{K-arcmin}$\\\hline\hline

220 GHz & 57 & 4.2&14.6&  $3.7\times 10^3$&435\\\hline
280 GHz  & 45  & 11.8 & 27.5&$2.0   \times 10^4$ &1140  \\\hline
350 GHz & 35 &85.1&105& $2.5\times 10^5$  &  5648.8    \\\hline
410 GHz & 30  & 468&377 &$6.6   \times 10^5$ &14174     \\\hline
850 GHz & 14 & 69483&575000&$1.5\times 10^7$&$  8.5\times10^6$\\\hline
\end{tabular}

\caption{Noise levels and beam sizes for the \Planck \cite{ refId0} , IRIS \cite{2005ApJS..157..302M}, and one of many possible CCAT-prime \cite{2020JLTP..199.1089C} configurations.  }\label{tab:noise_planck}
\end{center}
\end{table*}

\subsubsection{CMB Lensing noise specifications}

The lensing potential can be reconstructed from CMB temperature and polarization maps \cite{Okamoto:2003zw}; we consider various reconstruction noise scenarios including reconstruction noise in line with that of \Planck, with a Simons Observatory-like scenario (specifically the `goal' configuration from \cite{Ade:2018sbj}), and a Stage-4 (S4) like scenario such as \cite{Abazajian:2016yjj}. Additionally, we also compare to the case when there is no noise on the lensing reconstruction out to $L=1000$. The signal and noise for the various lensing scenarios are plotted on the right of Figure \ref{fig:ccatp_signal_and_noise}.  In all cases we assume that the CMB lensing potential is measured reliably in the multipole region $186\le L\le1000$.

With the signal and noise expected from these experiments, we can calculate the forecast errors on the measurement of the CIB/lensing cross correlation. The error-bars on $C_L^{\phi \nu}$ can be calculated from the bandpower covariance matrix of Equation \eqref{cov_bandpower};
with $\alpha=\gamma=\kappa$ and $\delta=\beta=\nu$ we can calculate the covariance of the $C_L^{\phi \nu}$
\ba{
\lb\Delta C_L^{\phi\nu}\rb^2=\frac{1}{\Delta L (2L+1)f_{\rm sky}}\bigg{[} \lb C_L^{\nu\nu}+N_L^{\nu\nu}\rb&\left(C_L^{\phi\phi}
+N_L^{\phi\phi}\right)
+\lb C_L^{\nu\phi}\rb^2\bigg{]}\label{error_cell_cross}
}
where $\Delta L$ is the width of the bins over which $C_L^{\nu\phi}$ is measured; in \cite{Ade:2013aro} are given in bins of width $\Delta L = 126$. See Figure \ref{fig:cross_errorbars} for plots of the predicted errors, the fiducial model, and the data of \cite{Ade:2013aro}.

\subsection{Foregrounds}\label{sec:foregrounds}

We include contributions from foregrounds as noise in the covariance matrices. The dominant foreground at high frequencies is emission from Galactic dust; however by restricting our baseline analysis to the cleanest 2240 sq. deg. used in \cite{2014A&A...571A..30P} and to angular scales with $L>186$, we substantially reduce the noise contribution from dust, and therefore do not include it in our forecasts.  The main contaminant at low frequencies is the CMB, which is dominant over the CIB at 217 and 353 GHz. Note, however, that for the analysis in \cite{2014A&A...571A..30P} the CMB was subtracted from these maps, using a template of the CMB measured at 143 GHz.  

We include the entire CMB power (computed with CAMB) in our forecasts at all frequencies, and find this has little effect on our forecast except for the predicted errors on the measured cross-power at CMB-dominated frequencies (see Figure \ref{fig:cross_errorbars}). We also include the early- and late-time kinetic Sunyaev--Zeldovich (kSZ) effect, the thermal Sunyaev--Zeldovich (tSZ) effect, and radio point sources as foregrounds \cite{2013JCAP...07..025D,2017PhRvD..96j3525M,Ade:2018sbj}.

\section{Forecast Results}\label{sec:fcresults}

\subsection{Constraints on CIB model parameters}\label{sec:parameterconstraints}
First we consider only \Planck+IRIS-like CIB data: $\nu=\{217, 353, 545, 857, 3000\}$ GHz, with noise specifications corresponding to those in Table \ref{tab:noise_planck}. For the CIB power spectra, we sum over all multipoles $186\le L\le2649$, corresponding to the multipoles used to fit the data in \cite{2014A&A...571A..30P}; for the CMB lensing power spectra we sum only over $186\le L\le1000$. We consider a sky fraction of 2240 square degrees for the \Planck frequencies $\nu=\{217,353,545,857\}$ GHz, and 183 square degrees for the 3000 GHz IRIS data, corresponding to the sky areas used in the \Planck analysis \cite{2014A&A...571A..30P}. We assume full overlap between all maps, and the CMB lensing potential reconstruction. Motivated by the priors used in \cite{2014A&A...571A..30P}, we include Gaussian priors in the Fisher matrix with widths of 20 K for $T_0$ and 0.5 for $\beta$; we also include Gaussian priors on the shot noise parameters with widths given by the $1\sigma$ errors in Tables 6 and 7 of \cite{2014A&A...571A..30P}. We assume the flux cuts in Table 1 of \cite{2014A&A...571A..30P}

 The forecast constraints on the parameters, and improvements upon including the different lensing scenarios, are given in Table \ref{tab:planck_constraints}. A bar graph of the improvement factors is also presented in Figure \ref{fig:bargraph}. Triangle plots of the covariances of the parameters are given in Figure \ref{fig:triangleplanck}. 

\begin{table*}[t]\label{tab:Planck_improvements}

\begin{tabular}{|c||c|c||c ||c||c||c|c||c|}
\hline
  & \multicolumn{2}{c||}{}&\multicolumn{1}{c||}{CIB-only: $N_L^{\nu\nu^\prime}{}_{\rm \it Planck} $ }&\multicolumn{5}{c|}{Forecast Improvement}\\\cline{2-9}
Parameter&\multicolumn{2}{c||}{Reported Constraint} &Forecast Constraint&  $N_L^{\kappa\kappa}{}_{\rm \it Planck} $& $N_L^{\kappa\kappa}{}_{\rm{SO}} $&  \multicolumn{2}{c||}{$N_L^{\kappa\kappa}{}_{S4} $} &  $N_L^{\kappa\kappa}=0$\\
\cline{2-9}
&Value&Percentage&$\sigma_{\rm CIB}$&${\sigma_{\rm CIB}}/{\sigma}$&${\sigma_{\rm CIB}}/{\sigma}$&${\sigma_{\rm CIB}}/{\sigma}$&Percentage&${\sigma_{\rm CIB}}/{\sigma}$\\\hline
$\alpha$ & 0.05 & 13.89 \% & 0.03 & 1.11 & 1.63 & 2.47 & 2.96 \% & 3.34 \\
$T_0$[K] & 1.9 & 7.79 \% & 1.04 & 1.22 & 1.89 & 2.46 & 1.73 \% & 2.88 \\
$\beta$ & 0.06 & 3.43 \% & 0.02 & 1.03 & 1.14 & 1.25 & 1.04 \% & 1.34 \\
$\gamma$ & 0.2 & 11.76 \% & 0.06 & 1.04 & 1.1 & 1.14 & 2.86 \% & 1.18 \\
$\delta$ & 0.2 & 5.56 \% & 0.29 & 1.17 & 1.59 & 1.85 & 4.42 \% & 2.02 \\
$\log_{10}(M_{\rm eff})$ & 0.1 & 0.79 \% & 0.17 & 1.08 & 1.2 & 1.28 & 1.05 \% & 1.37 \\
$L_0$ & None & & 0.0 & 1.19 & 1.62 & 1.87 & 12.45 \% & 2.04 \\
$\log_{10}(M_{\rm min})$ & unconstrained & & 47.81 & 1.06 & 1.38 & 1.87 & 255.99 \% & 2.36 \\
\hline
\end{tabular}
\caption{Constraints and improvement factors ${\sigma_{\rm CIB}}/{\sigma}$ on this model when only incorporating \Planck data. In the columns labelled `Percentage' we report the size of the constraint as a percentage of the fiducial parameter value.}\label{tab:planck_constraints}

\end{table*}

\begin{table*}[t]\label{tab:Planck_improvements}
\begin{tabular}{|c||c||c ||c||c|c||c|}
\hline
  &\multicolumn{1}{c||}{CIB-only: $N_L^{\nu\nu^\prime}{}_{\rm  CCAT-prime} $ }&\multicolumn{5}{c|}{Forecast Improvement}\\\cline{2-7}
Parameter&Forecast Constraint  &$N_L^{\kappa\kappa}{}_{\rm \it Planck} $& $N_L^{\kappa\kappa}{}_{\rm{SO}} $&  \multicolumn{2}{c||}{$N_L^{\kappa\kappa}{}_{S4} $} &  $N_L^{\kappa\kappa}=0$\\
\cline{2-7}
&$\sigma_{\rm{CIB}}$&${\sigma_{\rm CIB}}/{\sigma}$&${\sigma_{\rm CIB}}/{\sigma}$&${\sigma_{\rm CIB}}/{\sigma}$&Percentage&${\sigma_{\rm CIB}}/{\sigma}$\\\hline

\hline
  
$\alpha$ & 0.02 & 1.06 & 1.38 & 1.99 & 2.66 \% & 2.75 \\
$T_0$[K] & 0.72 & 1.1 & 1.54 & 2.06 & 1.42 \% & 2.52 \\
$\beta$ & 0.02 & 1.01 & 1.08 & 1.14 & 0.78 \% & 1.18 \\
$\gamma$ & 0.05 & 1.01 & 1.06 & 1.09 & 2.61 \% & 1.12 \\
$\delta$ & 0.22 & 1.1 & 1.52 & 1.96 & 3.12 \% & 2.29 \\
$\log_{10}(M_{\rm eff})$ & 0.13 & 1.08 & 1.34 & 1.55 & 0.67 \% & 1.69 \\
$L_0$ & 0.0 & 1.1 & 1.53 & 1.95 & 8.84 \% & 2.26 \\
$\log_{10}(M_{\rm min})$ & 37.67 & 1.05 & 1.33 & 1.88 & 200.05 \% & 2.58 \\\hline
\end{tabular}
\caption{$1\sigma$ constraints and improvement factors on the parameters when including both \Planck and CCAT-prime data.}\label{tab:ccatp_constraints}
\end{table*}

Second we consider CCAT-prime+IRIS-like data. As CCAT-prime is noise dominated at low $L$, we include the low-$L$ data from \Planck in this forecast as well. Thus we consider a forecast at $\nu= \{220,  280,  350,  410, 545, 850, 3000\}$ GHz, although the 545 and 3000 GHz data is only signal-dominated at low $L$, and the 410 and 280 GHz data are only signal dominated at high $L$. At 220, 350, and 850 GHz respectively we consider \Planck noise levels appropriate to 217, 353, and 857 GHz.

We must also include shot noise values in the CCAT-prime forecast. We consider flux cuts similar to the \Planck experiment; as the flux cuts could be smaller and the shot noises lower, our forecast is conservative in this regard. We choose the shot noise parameters from those in Table \ref{tab:shot_noises}, where we have in every case rounded up the relevant CCAT-prime frequency if it does not appear in the table. 
Forecast 1$\sigma$ constraints are shown in Table \ref{tab:ccatp_constraints}. 

\subsection{Impact of the high-frequency data}\label{sec:3000}

The forecasts in Sec. \ref{sec:fcresults} included 3000 GHz data on around 8\% of the sky area on which the low-frequency fields are measured. The 3000 GHz field is qualitatively different to the low frequency fields: it probes the high-frequency end of the SED \eqref{SED} and as such informs the parameter $\gamma$ while the lower frequency fields do not. It is also sourced at lower redshift (see Figure \ref{fig:redshift_distribution}) and thus is less correlated with the rest of the CIB and can provide more independent information. As well as providing all the information on $\gamma$, including the 3000 GHz field  on even this small sky fraction provides significant constraining power on the parameters relating to the dust temperature ($\alpha$, $T_0$), improving the constraints on these parameters by up to 100\%. Because of this, it is important to include the high-frequency information in any CIB model fitting.

\subsection{Galaxies as an external tracer}\label{sec:galaxies}
The Rubin Observatory \cite{2009arXiv0912.0201L} will measure the clustering of billions of galaxies and their photometric redshifts in their LSST (Legacy Survey of Space and Time) survey. While the CMB lensing kernel is highly correlated with the CIB, both peaking at redshift $z\sim2$, it is interesting to see what low-redshift information can add to the CIB, particularly as the 3000 GHz field (which is sourced at lower redshift) helps significantly with some parameters (as discussed in Sec. \ref{sec:3000}). In this section we consider how a low-redshift galaxy sample from the Rubin Observatory can help improve parameter constraints. A similar analysis was done in \cite{Serra:2014pva} where a CIB halo model was fit to the cross power spectra of the CIB and SDSS galaxies in a narrow redshift bin, in order to isolate redshift behaviour of the CIB.

The angular galaxy clustering power-spectrum for a photometric redshift bin between redshifts $z_i$ and $z_f$ (comoving distances $\chi_i$ and $\chi_f$) 
\be
C_L^{gg}= \frac{1}{\Delta \chi^2}\int_{\chi_i}^{\chi_f}\frac{ d\chi} {\chi^2}P_{gg}\lb k=\frac{L}{\chi},z\rb,
\ee
where $\Delta \chi=\chi_f-\chi_i$ is the extent of the bin in comoving distance and $P^{gg}$ is the galaxy power spectrum. We choose to consider only one photometric redshift bin, from $z=0$ to $z=1$, and use only the two-halo galaxy power spectrum (Equation \eqref{gal2halo}), which can be written as
\be 
P_{gg}^{\rm 2-halo}(k,z)=b_g^2P_{\rm lin}(k,z),
\ee
where the galaxy bias $b_g$ is defined as
\be
b_g(z)= \int dM \frac{dN}{dM}\frac{N^{\rm gal}(M,z)}{\bar n_{\rm gal}(z)} b(M,z).\label{bias}
\ee
For the galaxy density field predicted for the LSST Gold sample \cite{2009arXiv0912.0201L}
\be
\frac{dn}{dz} \propto z^2\exp{\lb-\frac{z}{0.5}\rb}\label{gold}
\ee
with a total number density of 40 $\text{arcmin}^{-2}$. This can be used to compute the total angular number density of galaxies and can be related to the number density of galaxies in the halo model given in \eqref{ngal}. Equation \eqref{ngal} gives the total number density of galaxies at $z$; only the most luminous (or massive) are seen by the galaxy survey and so specifying a galaxy distribution \eqref{gold} is equivalent to specifying a $z$-dependent minimum mass in the integral \eqref{ngal}. 

As we do not wish to focus on uncertainties in the non-linear galaxy HOD, we restrict our galaxy clustering information to scales where only the two-halo term is relevant by using an $L_{max}$ of 500 for the galaxy survey and neglecting the 1-halo terms. 

We can write the two-halo cross-power spectrum between galaxies and the CIB emissivity as 
\be
\bar j (z)P^\nu_{j g}{}^{\rm 2-halo}(k,z)=b_g(z)D_\nu(z)P_{\rm lin}(k,z)
\ee
with $D_\nu(z)$ the CIB bias. The angular power spectra can be computed from the Limber approximation
\be
C_L^{\nu g}=\frac{1}{\Delta \chi}\int _{\chi_i}^{\chi_f}\frac{d\chi}{\chi^2}  j (z)P^\nu_{j g}\lb k=\frac{L}{\chi},z\rb.
\ee
We perform a forecast with the same formalism as in Sec. \ref{sec:fisher} where we now consider a covariance matrix with galaxy clustering included:
\be
C_L=\left(
\begin{array}{c c c}
C_L^{\nu\nu^\prime}&C_L^{\nu\phi} &C_L^{\nu g} \\
C_L^{\nu\phi}&C_L^{\phi\phi} & C_L^{\phi g}\\
C_L^{\nu g} & C_L^{\phi g} & C_L^{gg}
\end{array}
\right).
\ee
The cross-power spectra between lensing and galaxies $C_L^{g\phi}$ is
\be
C_L^{\phi g}=\frac{1}{\Delta \chi l^2}\int \frac{d\chi}{\chi^2} a(\chi) b_g(\chi) W_\kappa(\chi) P_{\rm lin}\lb k=\frac{L}{\chi},z\rb.
\ee
To account for the uncertainties in the modelling of our galaxy power spectra, when we include galaxies in the forecast we marginalise over the galaxy bias by introducing a parameter $A$ such that $C_L^{gg}=A^2C_L^{gg}$,  $C_L^{g\phi}=AC_L^{g\phi}$, and $C_L^{g\nu}=AC_L^{g\nu}$ with $A=1$ in the fiducial case.

We assume full overlap between the galaxy field, the CIB fields, and the CMB lensing field (although we restrict the 3000 GHz field to 183 square degrees as before). We find that some CIB model parameters can be constrained much more strongly when including galaxies;  similarly to when the 3000 GHz field was included, the CIB dust temperature parameters  $\alpha$ and $T_0$ are improved significantly, as well as the parameter controlling the redshift evolution of the $L-M$ normalisation $\delta$, indicating that the low-redshift information helps to inform these parameters.

\section{Constraints on star formation history}\label{sec:sfr}

The source of the energy of the dust particles emitting the CIB is irradiation by ultraviolet (UV) light emitted by the star-forming galaxies. The star formation rate (SFR) can be measured directly with UV detections; however these measurements must be corrected for the dust attenuation, as much of the UV emission is indeed absorbed and re-emitted in the IR (see \cite{doi:10.1146/annurev-astro-081811-125615} for a review of cosmic star formation history). In \cite{2012A&A...539A..31C}, UV measurements are used to constrain the SFRD at redshifts up to $z=4.5$. Direct measurements of total IR emission of galaxies are also used to constrain SFR \cite{magnelli_SFR,gruppioni_SFR}. As the CIB emission traces all galaxies (not just those luminous enough to be resolved as sources), it can provide a complementary probe of the SFR, particularly at high redshift.

The star formation rate (SFR) can be related very simply to the total infrared luminosity of galaxies through the Kennicutt relation \cite{1998ARA&A..36..189K}:
\be
\mathrm{\rm SFR} = K L_{\rm IR}
\ee
with $K$ the Kennicutt constant $K = 1.7\times 10^{-10} M_\odot \mathrm{yr}^{-1}$. The total infrared luminosity is simply the  luminosity density integrated over its entire IR emission spectrum:
\be
L_{\rm IR} = \int d\nu L_\nu
\ee
---as the only $\nu$-dependence is in the SED this can be written equivalently as
\be
L_\nu = \Theta_\nu L_{\rm IR}
\ee
with the SED $\Theta(\nu)$ normalised such that $\int d\nu \,\Theta(\nu)=1$\footnote{Note that this is a different normalisation to the SED in Section \ref{sec:IR_SED}.}. The definition of emissivity \eqref{j_def} can then be written
\be
 j_\nu(z)=\Theta_{(1+z)\nu}\int dL_{\rm IR}\frac{d N}{dL_{\rm IR}}\frac{ L_{\rm IR}}{4\pi}\label{jnu_LIR}
\ee
with $\frac{d N}{dL_{\rm IR}}$ the IR luminosity function such that $\frac{d N}{dL_{\rm IR}} dL_{\rm IR}$ gives the number density of halos with total IR luminosity between $L_{\rm IR}$ and $L_{\rm IR} + dL_{\rm IR}$. Due to the Kennicutt relation, the integral in \eqref{jnu_LIR} gives the mean star formation rate density (SFRD) $\rho_{\rm SFR}$:
\be
 j_\nu(z)=\frac{\Theta_{(1+z)\nu}}{4\pi}\frac{ \rho_{\rm SFR}(z)}{K}.
\ee
This can be written in terms of the effective SED $s_{\nu,\rm eff}$, the flux density from a halo with luminosity of $1 L_{\odot}$ 
\be
s_{\nu,\rm eff} =  \Theta_{(1+z)\nu}\frac{1 L_{\odot}}{4\pi\chi^2(1+z) }\label{snueff}
\ee
such that
\be
j_\nu(z) = \frac{\rho_{\rm SFR}(z)s_{\nu,\rm eff}\chi^2 (1+z)}{K}.
\ee
In modelling the star formation rate, an alternative approach to using a parametric SED is to use for $s_{\nu,\rm eff}$ externally measured SEDs such as those of \cite{Bethermin_SEDs}. Indeed, in \cite{Maniyar:2018xfk} CIB and CMB lensing data are used to constrain the $\rho_{\rm SFR}$ in this way. This approach has the advantages of being able to incorporate different types of galaxies with different SEDs \cite{Bethermin_SFR} such as those undergoing a starburst phase or the more common main sequence galaxies.

We can use the parametric halo model to compute $\rho_{\rm SFR}(z)$ by using for the parametric SED of Equation \eqref{SED_parametric} in \eqref{snueff} to compute $s_{\nu,\rm eff}$ (note it must be normalised to integrate to $1$ over all frequencies). We can then forecast the constraints on $\rho_{\rm SFR}(z)$ by drawing parameters from the covariance matrix defined by $F^{-1}$, the inverse of the Fisher matrices discussed above. We show in Figure \ref{fig:sfr} how the inclusion of lensing data improve constraints on $\rho_{\rm SFR}$ through this model; as it is difficult to see the improvements on a logarithmic scale we include a linear of the constraints divided by the fiducial value of $\rho_{\rm SFR}$. We define the $1\sigma$ errors as the area within which 68\% of 1000 realisations fell, centered on the median value.

\section{Discussion}\label{sec:conclusion}

In this work, we explored the possibility of using cross-correlations of CMB lensing mass maps with maps of the CIB to improve physical models of the latter. We have shown that inclusion of CMB lensing data can lead to up to $2\times$ improvement in constraints on the dust temperature and its redshift evolution, and on the redshift evolution of the relation between CIB galaxy luminosity and mass, in particular. Since cosmological parameters like the amplitude of matter fluctuations are known to much better precision than the astrophysical parameters of interest here, we have not varied them in our forecasts (although see \cite{Reischke:2019ikn} for a CIB-only forecast which varies the cosmological parameters along with the CIB parameters). Therefore, the CMB lensing potential does not depend on the parameters in consideration. Due to this, improvements in parameter constraints will come from either the redshift overlap or cancellation of sample variance, as described below.

Figure \ref{fig:planck_signal_and_noise} shows that at 353, 545 and 857 GHz, the \Planck instrument noise is at least an order of magnitude lower than the CIB signal itself over a wide range of scales, typically extending out to $L>1000$. The two-halo CIB signal is thus mostly limited by sample variance in the matter density traced by the CIB, and parameters of the halo model we are interested in will not improve with improved noise for a given frequency configuration, though see below for a discussion on the importance of higher frequencies. When parameters of a model are limited by sample variance in some field, say, $\delta_g = b \delta_m$ (where $b$ is the parameter of interest), further improvements can however be obtained through joint measurement with an additional field that traces the same underlying fluctuations (see e.g. \cite{Seljak:2008xr,Schmittfull:2017ffw,Munchmeyer:2018eey}). For example, if one were to measure the $\hat{\delta}_m$ field itself in addition to $\hat{\delta}_g$, the uncertainty on $\hat{b}=\hat{\delta}_g / \hat{\delta}_m$  will not depend on the fluctuations in $\hat{\delta}_m$ due to cancellation of the sample variance. In practice, one needs to compute all auto and cross-spectra between the relevant fields and incorporate it in the likelihood for the parameters of interest. The gains from sample variance cancellation can be substantial if the cross-correlation coefficients between the two fields are close to unity. In our work, we use the projected mass density measured by CMB lensing, which has significant redshift overlap with the CIB (Figure \ref{fig:redshift_distribution}), which leads to correlation coefficients larger than 70\% out to $L=800$ even in the presence of instrument noise and foregrounds in the CIB for perfect reconstruction of the lensing field and out to $L=400$ for CMB-S4 levels of noise in the lensing reconstruction (see Figure \ref{fig:corrcoeff}).

We have calculated explicitly the constraints for a CIB-only scenario with zero foregrounds and instrumental noise, and we find that in the realistic forecast,  while not all of the parameters have reached their sample variance limit, for many of them the inclusion of SO- or S4-like lensing improves the constraint to such an extent that they are better than the ``perfect'' sample-variance limited CIB-only case. However, we note that this effect might not be due solely to sample variance cancellation, but that the well-understood redshift kernel of the CMB lensing field may be helping to constrain the redshift dependence of the CIB fields in a similar manner.

In order to be conservative and facilitate comparisons with earlier work, the improvements we have presented are calculated for the small sky areas ($\sim5\%$) used for the analysis in \cite{2014A&A...571A..30P}. It is possible to consider larger sky areas for the CIB maps than was done here; \cite{Lenz:2019ugy}  produced maps with improved treatment of galactic dust using HI data and recommended sky fractions of \{18.7\%, 16.3\%, 14.4\%\} at \{353, 545, 857\}\ GHz for auto-power spectrum analysis. Accounting for the partial sky covered by typical ground-based surveys however reduces the area available for cross-correlation (necessary for improvements from sample variance cancellation) to  \{6.4\%, 8.4\%, 9.8\%\}, respectively. 

\begin{figure*}[t]
\includegraphics[width=0.99\textwidth]{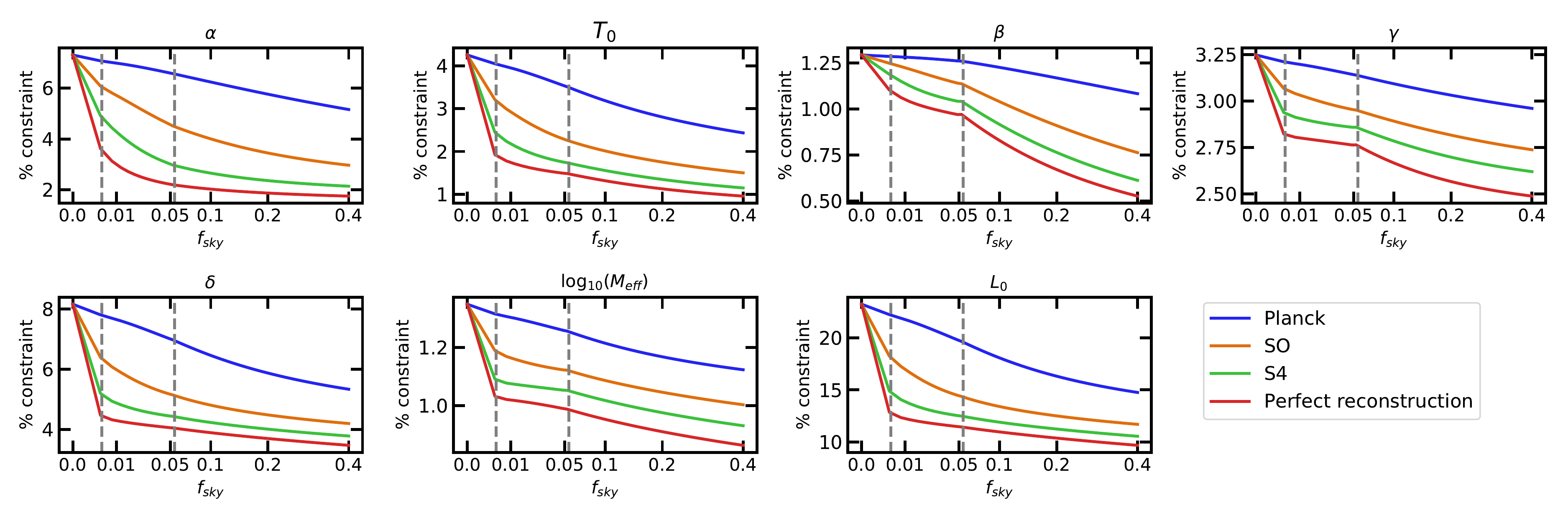}
\caption{The improvement in CIB model constraints with the area of the CIB-lensing cross correlation included in the analysis. In all cases the CIB auto power spectrum is measured on the fiducial 2240 square degrees (183 square degrees for 3000 GHz), and the lensing field is introduced first on the small 183 square degree patch where we have the entire \Planck+IRIS fields, then onto the 2240 square degrees where we have the \Planck fields, and finally it is included without the CIB auto power spectra measured on the same patch. The two dashed grey lines denote these transitions. The $x$-axis has been transformed such that it is linear in $\sqrt{f_{\rm sky}}$.}\label{fig:area}
\end{figure*}

The requirements on foreground cleaning for the CIB maps are more stringent for an auto-power-spectrum analysis than for a cross-correlation with CMB lensing. Much of the foreground contamination is sourced by Galactic dust, which will introduce spurious correlations in the auto-spectrum from the spatially dependent two-point correlation of Galactic emission, which can be significantly brighter than the CIB. In a cross-correlation of CMB lensing (calculated through quadratic estimators of the form $\langle T^{150} T^{150}\rangle$, where $T^{150}$ is the CMB temperature field as measured at 150 GHz that dominates near-term experiments) with the CIB, on the other hand, biases enter through bispectra of the form $\langle T^{150}_G T^{150}_G T^{\rm high}_G \rangle$ where $T^{150}_G$ is the Galactic dust emission at 150 GHz and $T^{\rm high}_G$ is the Galactic emission at high frequencies used for CIB maps. These are suppressed relative to the biases in the CIB auto-spectrum for several reasons that include (1) the SED of dust being such that $T^{150}_G$ is significantly smaller than $T^{\rm high}_G$ and (2) contributions from Galactic dust blobs being further reduced in the high-resolution CMB map through mitigation techniques like point source bias hardening \cite{Osborne}.  As such, the sky area available for the cross-correlation between the CIB and CMB lensing is larger; the largest maps of \cite{Lenz:2019ugy} have a total sky area of 34.2\% with roughly 20\% overlap with typical wide-area ground based high-resolution CMB experiments from which CMB lensing maps will be available. To understand the impact of sky area, we show in Figure \ref{fig:area} the improvement in the constraints as the lensing field is added on top of the baseline forecast. The area on which the CIB auto power spectrum is measured is not changed in Figure \ref{fig:area}, and the lensing field is added first on the IRIS+\Planck fields, then on the \Planck fields, and then on extra sky but without any CIB auto-power spectrum measured beyond the baseline 2240 sq. deg. We find that for some parameters like the SED emissivity index $\beta$, substantial improvements can be obtained by including CIB/lensing cross-correlations in larger fractions of the sky, without having to measure the CIB auto-spectrum beyond 2240 sq. deg. We note however that our forecasts do not include the increased scatter on large scales from Galactic dust contamination that is encountered when including larger sky areas \cite{Lenz:2019ugy}. 

We have checked that forecasts with the auto-spectrum of the CMB lensing map removed---i.e., with only $C_L^{\nu\nu^\prime}$ and $C_L^{\nu\kappa}$ as data---and have found that there is negligible degradation in the forecast parameter improvements. As such, one will not need to measure the auto-spectrum of the lensing potential to achieve model improvements.

We find that the inclusion of the 3000 GHz field, even on a small fraction of the sky area, is key for constraining the CIB model. This is not only due to the fact that it is the only field which informs the high-frequency part of the SED, but also because it is the only low-redshift tracer included in the survey, as evidenced in Figure \ref{fig:redshift_distribution}. As such, it provides information that is independent of the low-frequency fields. We have also demonstrated that the inclusion of another low-redshift tracer (such as galaxy clustering out to $z\sim1$, say from the Rubin Observatory \cite{2009arXiv0912.0201L,desc}) can improve parameters even further. As in the case for when lensing is included, the galaxy density must be measured on the same patch of sky to provide improvement. This is an interesting possibility, however we note that we have presented rough optimistic forecasts as an illustrative example of the power of correlating the CIB with other low $z$ tracers to improve the CIB model.

For robustness, our baseline forecasts have restricted analyses  mostly to the mostly two-halo regime by using the multipole range $186\le L\le2649$. We have however also considered improvements to a CIB model made with CIB maps of higher angular resolution exploring the one-halo regime, in particular those that will be made by CCAT-prime. While the HOD we use is not expected to be very accurate on such small scales, it is interesting that there is still improvement factors up to $\sim 2$ for the S4 case in these forecasts when the large-scale lensing cross-correlation is included. 

Studying the CIB is interesting for a variety of reasons. Firsty, from an astrophysical perspective, it contains interesting information about star formation history. We have shown what improvements there can be made to measurements of the star formation rate density through including lensing data in Sec. \ref{sec:sfr}. While the model we use as our fiducial model may not reproduce very accurately other measurements of the star formation rate (perhaps due to the crude parametric SED used), it is possible that alternative models of the CIB, such as the simpler single-parameter halo model used in \cite{Maniyar:2020tzw} (which uses externally measured SEDs) will be improved similarly by including lensing data.  The CIB is also an interesting cosmological signal: as a tracer of large scale structure, the CIB carries interesting information \cite{Maniyar:2018hfp,McCarthy:2019xwk} and having a more accurate model can allow us to exploit further its cosmological information.

Secondly, the CIB is important to understand as a foreground to other signals of interest. The CIB is a significant foreground to the CMB at small angular scales and its accurate modelling is necessary to make unbiased measurements of signals that are relevant to the small-scale CMB such as the kSZ power-spectrum \cite{Reichardt:2020jrr}, as well as extensions to the $\Lambda$CDM model deriving significant information from the damping tail\footnote{It should be noted that as the polarization sensitivity of ground-based experiments improves, parameters like the number of relativistic species $N_{\rm eff}$ will increasingly derive their information from the polarization TE and EE spectra. Thus, CIB-related model bias to extensions of the $\Lambda$CDM model (like $N_{\rm eff}$) will likely be less of an issue, since the CIB is not significantly polarised. }. We leave detailed exploration of the potential improvements to physics in the temperature damping tail to future work. The tSZ effect is also a significant foreground to the CMB at small angular resolution, and must also be mitigated or modeled; as discussed in \cite{Maniyar:2020tzw}, the tSZ/CIB correlations can be consistently modelled with the halo model approach presented in Sec. \ref{sec:halo_model}. As we find that significantly more accurate models of the CIB can be built by including external tracers, in particular the CMB lensing potential, in the data analysis, this method of constraining models of the CIB will be of great use in improving our knowledge of star formation as well as potentially physics in the damping tail.

\begin{acknowledgements}
We thank Abhishek Maniyar, Robert Reischke, and Guilaine Lagache for very useful correspondence in particular regarding the normalisation of the CIB SED.  We also thank Niall MacCrann, Yogesh Mehta, Colin Hill, Neelima Sehgal, Blake Sherwin and Alexander van Engelen for useful discussions. We thank Steve Choi and Nick Battaglia for information on the CCAT-prime survey.  FMcC acknowledges support from the Vanier Canada Graduate Scholarships program. Research at Perimeter Institute is supported in part by the Government of Canada through the Department of Innovation, Science and Economic Development Canada and by the Province of Ontario through the Ministry of Colleges and Universities.
\end{acknowledgements}

\appendix

\section{Conversion between $\mathrm{\mu K}$ and $\mathrm{Jy}$}\label{App:units}

In this paper we have presented the CIB power spectra in $\mathrm{Jy}$, a unit of surface intensity commonly used for CIB measurements and in radio astronomy. However, we quote the CMB lensing power spectra in $\mathrm{\mu K_{CMB}}$; additionally, some of the instrumental noise levels we quote are in $\mathrm{\mu K}$. Thus, as it is convenient to have a formula to convert between these units $\mathrm{Jy}$ and $\mathrm{\mu K_{CMB}}$, we present one in this Appendix.

\begin{table}[h!]
\begin{tabular}{|c|c|}
\hline
$\nu[\mathrm{Ghz}]$&$U\,[\mathrm{Jy\,\mu K_{CMB}^{-1}}]$\\\hline\hline
217 & 483.69\\\hline
353 &287.45\\\hline
545 &58.04 \\\hline
857 &2.27 \\
\hline
\end{tabular}
\caption{Conversion factors between $\mathrm{Jy}$ and $\mu K_{CMB}$, from \cite{2014A&A...571A...9P}.}\label{tab:example_convert}
\end{table}

The surface brightness of a black body is given by the Planck formula
\be
B_\nu(T) = \frac{2h\nu^3}{c^2}\frac{1}{e^{\frac{h\nu}{kT}}-1}.
\ee
To convert from brightness to temperature we use
\be
dB_\nu(T)=\frac{2 h^2 \nu ^4 e^{\frac{h \nu }{k T}}}{c^2 k T^2 \left(e^{\frac{h \nu }{k T}}-1\right)^2} dT.
\ee
Defining 
\be
x\equiv \frac{h \nu }{k T_{\rm CMB}} = \frac{\nu [\mathrm{GHz}]}{56.233 \mathrm{GHz}}
\ee
and using the definition of a Jansky $\mathrm {Jy} = 10^{-26} \frac{\mathrm{W}}{\mathrm{m}^2\mathrm{Hz}}$ we can write
\be
dB_\nu [\mathrm{MJy}]= 968 \frac{e^x\lb\frac{\nu [\mathrm{GHz}]}{100}\rb^4}{\lb e^x-1\rb^2} \mathrm{\mu K}.
\ee

While this formula is useful, in general a more accurate conversion between the units is dependent on the specifications (spectral response, etc) of the instrument used and so for \Planck frequencies we use the units quoted in \cite{2014A&A...571A...9P}; see Table \ref{tab:example_convert}.

\bibliography{references}

\end{document}